%% file: main.tex
\documentclass[a4paper]{article}

\usepackage[left=2.25cm, right=2.25cm, top=2cm, bottom=2cm, marginparwidth=1.7cm, marginparsep=2mm]{geometry}
\usepackage[utf8]{inputenc}
\usepackage[T1]{fontenc}
\usepackage{cmbright}
\usepackage{graphicx}
\usepackage[table, dvipsnames]{xcolor}
\usepackage{amsfonts}
\usepackage{bm}
\usepackage{amsmath}
\usepackage{amsthm}
\usepackage{mathtools, nccmath}
\usepackage{url}
\usepackage{booktabs}
\usepackage{tabularx}
\usepackage{nicefrac}
\usepackage{microtype}
\usepackage[hidelinks]{hyperref}
\hypersetup{
    colorlinks=true,
    linkcolor=blue,
    urlcolor=blue,
    citecolor=blue,
    anchorcolor=blue}
\usepackage[inline]{enumitem}
\usepackage{float}
\usepackage{wrapfig}
\usepackage[font=footnotesize, labelfont=bf]{caption}
\usepackage{subcaption}
\usepackage{adjustbox}
\usepackage{tikz}
\usetikzlibrary{shapes.geometric}
\input{boxes.tex}

\usepackage{changepage}
\usepackage{etoc}
\usepackage{ragged2e}
\usepackage[framemethod=TikZ]{mdframed}
\usepackage[super, comma, numbers]{natbib}
\usepackage[symbol]{footmisc}
\usepackage{multicol}
\usepackage{lipsum}
\usepackage{orcidlink}
\definecolor{orcidlogocol}{HTML}{000000}
\usepackage{xspace}
\usepackage{multirow}

\definecolor{darkblue}{HTML}{1A254B}
\definecolor{lightblue}{HTML}{A7BED3}
\definecolor{blue}{HTML}{114083}
\definecolor{green}{HTML}{81B5AE}
\definecolor{pink}{HTML}{F2545B}
\definecolor{red}{HTML}{A4243B}
\definecolor{lightgray}{HTML}{C3BABA}
\definecolor{darkgray}{HTML}{9A8F97}

\title{\vspace{-30pt} \bf AI-powered virtual tissues from spatial proteomics for clinical diagnostics and biomedical discovery}
\date{}

\definecolor{hanpurple}{rgb}{0.32, 0.09, 0.98}

\newcommand{\our}{\text{VirTues}\xspace}
\newcommand{\camae}{\text{CA-MAE}\citep{kraus2024masked}\xspace}
\newcommand{\resnet}{\text{ResNet}\citep{sorin2023single}\xspace}
\newcommand{\kronos}{\text{KRONOS}\citep{shaban2025foundation}\xspace}

\newcommand{\danenberg}{\citet{danenberg2022breast}\xspace}

\newcommand{\hoch}{\citet{hoch2022multiplexed}\xspace}
\newcommand{\cords}{\citet{cords2024cancer}\xspace}

\newcommand{\R}{\mathbb{R}}
\newcommand{\ceil}[1]{\left\lceil #1 \right\rceil}

\newcommand{\setconstrain}[2]{\left\{#1 \, \mid \, #2\right\}}
\newcommand{\norm}[1]{\left\lVert#1\right\rVert}
\newcommand{\set}[1]{\left\{#1\right\}}

\newlength{\wdth}

\usepackage{xr}
\usepackage{lineno}

\author{%
\small Johann Wenckstern$^{\orcidlink{0009-0007-1511-4659}\,1,9}$\,, 
\small Eeshaan Jain$^{\orcidlink{0009-0002-6041-7193}\,1,9}$\,, 
\small Yexiang Cheng$^{\orcidlink{0009-0005-2407-6653}\,1,9}$,
\small Benedikt von Querfurth$^{\orcidlink{0009-0001-0384-7923}\,1,2}$,\\
\small Kiril Vasilev$^{3}$,
\small Matteo Pariset$^{3}$,
\small Phil F. Cheng$^{\orcidlink{0000-0003-2940-006X}\,4,5}$,
\small Petros Liakopoulos$^{\orcidlink{0009-0005-2015-6795}\,4,5}$,\\
\small Olivier Michielin$^{\orcidlink{0000-0003-4926-6355}\,4,5}$,
\small Andreas Wicki$^{\orcidlink{0000-0002-2924-8080}\,6,7}$\,,
\small Gabriele Gut$^{\orcidlink{0000-0001-8991-0040}\,6,7}$\,,
\small Charlotte Bunne$^{\orcidlink{0000-0003-1431-103X}\,1,8,10}$\\
}

\begin{document}

\maketitle

\vspace{-10pt}
{\scriptsize
\noindent $^{1}$ School of Computer and Communication Sciences, EPFL, Lausanne, Switzerland, \\
$^{2}$ Faculty of Computer Science, RWTH Aachen University, Aachen, Germany\\
$^{3}$ Department of Computer Science, ETH, Zurich, Switzerland, \\
$^{4}$ Faculty of Medicine, University of Geneva, Geneva, Switzerland,\\
$^{5}$ Department of Oncology, Geneva University Hospitals, Geneva, Switzerland,\\
$^{6}$ Department of Medical Oncology and Hematology, University Hospital Zurich, Zurich, Switzerland, \\  
$^{7}$ University of Zurich, Faculty of Medicine, Zurich, Switzerland, \\
$^{8}$ Swiss Institute for Experimental Cancer Research, School of Life Sciences, EPFL, Lausanne, Switzerland. \\
$^{9}$ These authors contributed equally. \\
$^{10}$ Correspondence to: \href{mailto:charlotte.bunne@epfl.ch}{charlotte.bunne@epfl.ch}. \\
}

\begin{multicols}{1}

\input{content/abstract.tex}

\input{content/introduction.tex}
\input{content/results.tex}
\input{content/discussion.tex}

{\footnotesize
\bibliographystyle{unsrtnat}
\bibliography{references}
}

\input{content/methods.tex}

\section*{Data availability}
All datasets used in this study are openly available from public sources with the exception of the data from \citet{rigamonti2024integrating}, which is available from the respective authors upon request. The multiplex images of \citet{allam2022spatially}, \citet{cords2023cancer}, \citet{cords2024cancer}, \citet{danenberg2022breast},
\citet{hoch2022multiplexed}, \citet{hu2023single}, \citet{schulz2024immucanpanel1}, \citet{jackson2020single}, \citet{moldoveanu2022spatially}, \citet{wang2023spatial} and \citet{meyer2025stratification} are accessible via Zenodo (\url{https://zenodo.org/records/6784253}, \url{https://zenodo.org/records/7540604}, \url{https://zenodo.org/records/7961844}, \url{https://zenodo.org/records/5850952}, \url{https://zenodo.org/records/6004986}, \url{https://zenodo.org/records/6784251}, \url{https://zenodo.org/records/12912567}, \url{https://zenodo.org/records/3518284}, \url{https://zenodo.org/records/5903190}, \url{https://zenodo.org/records/7990870} and \url{https://zenodo.org/records/10942609}). \citet{damond2019map} and \citet{schulz2018simultaneous} are downloadable from Mendeley Data (\url{https://data.mendeley.com/datasets/cydmwsfztj/1} and \url{https://data.mendeley.com/datasets/m4b97v7myb/1}). The dataset of \citet{zhu2025spatial} can be obtained from Synapse (\url{https://www.synapse.org/Synapse:syn54951674/wiki/626860}). Sequences of the protein markers were obtained from UniProtKB (\url{https://www.uniprot.org/}).
\section*{Code availability}
All code was implemented in Python using PyTorch as the primary deep learning package. The source code for the model and Supplementary Information for \our\ is available at \url{http://github.com/bunnelab/virtues}. 

\section*{Acknowledgments}
We thank Bernd Bodenmiller, Eric Lubeck, Zoe Piran, Lucas Pelkmans, Lukas Klein, Linus Bleistein and Aviv Regev for discussions and for providing feedback on our manuscript. We acknowledge support from Lena Cords, Daniel Schulz, Federica Marchesi, Marika Viatore and Lasse Meyer with accessing and analyzing the datasets.
We are very grateful for Andreas Krause's support and hosting J.W., K.V. and C.B. during the initial phase of this study. Icons were created using resources from Flaticon.com.

\section*{Author contributions.}
J.W., G.G. and C.B. conceived the study; J.W., Y.C., M.P., G.G. and C.B. devised the model architecture and its pretraining; J.W., E.J., Y.C., B.v.Q., K.V., P.F.C. and P.L. curated the datasets; J.W., E.J., Y.C., B.v.Q., K.V., G.G. and C.B. developed the evaluation framework and downstream tasks; J.W., E.J., Y.C., B.v.Q. and K.V. performed the experiments; J.W., E.J., A.W., G.G. and C.B. wrote the manuscript; O.M., A.W. and C.B. funded the study. All authors approved the final version of the manuscript.

\section*{Declaration of interests}
C.B. serves on the scientific advisory board of Ensocell Therapeutics. No patent applications have been filed on this work.
\end{multicols}

\newpage
\setcounter{page}{1}

\setcounter{section}{0}
\renewcommand{\thesection}{\Alph{section}}
\setcounter{table}{0}
\renewcommand{\tablename}{Table}
\renewcommand{\thetable}{S\arabic{table}}%
\setcounter{figure}{0}
\renewcommand{\figurename}{Figure}
\renewcommand{\thefigure}{S\arabic{figure}}%

\end{document}


\appendix
\section*{\LARGE Supplementary Material}

\section*{Supplementary Figures}

\begin{figure*}[h!]
    \centering
    \includegraphics[width=\textwidth]{figures/supplement/legend_datasets.pdf}
    \caption{Color legend for the 15 datasets used in this study.}
    \label{suppfig:legend_datasets}
\end{figure*}

\begin{figure*}[h!]
    \centering
    \includegraphics[width=\textwidth]{figures/supplement/examples_independent_1.pdf}
    \caption{Examples of image reconstruction results using independent masking, with masking ratios between 60\% and 100\%, for \cords, \damond, \danenberg, \hoch and \jackson.}
    \label{suppfig:examples_independent_1}
\end{figure*}

\begin{figure*}[h!]
    \centering
    \includegraphics[width=\textwidth]{figures/supplement/examples_independent_2.pdf}
    \caption{Examples of image reconstruction results using independent masking, with masking ratios between 60\% and 100\%, for \citet{zhu2025spatial}, \citet{schulz2024immucanpanel1}, \citet{rigamonti2024integrating}, \citet{hu2023single} and \citet{cords2023cancer}.}
    \label{suppfig:examples_independent_2}
\end{figure*}

\begin{figure*}[h!]
    \centering
    \includegraphics[width=\textwidth]{figures/supplement/examples_independent_3.pdf}
    \caption{Examples of image reconstruction results using independent masking, with masking ratios between 60\% and 100\%, for \citet{moldoveanu2022spatially}, \citet{schulz2018simultaneous}, \citet{allam2022spatially} and \citet{wang2023spatial}.}
    \label{suppfig:examples_independent_3}
\end{figure*}

\begin{figure*}[h!]
    \centering
    \includegraphics[width=\textwidth]{figures/supplement/examples_marker_1.pdf}
    \caption{Examples of image reconstruction results after masking an entire marker for \cords, \damond, \danenberg, \hoch and \jackson. Each row depicts different channels of the same tissue niche.}
    \label{suppfig:examples_marker_1}
\end{figure*}

\begin{figure*}[h!]
    \centering
    \includegraphics[width=\textwidth]{figures/supplement/examples_marker_2.pdf}
    \caption{Examples of image reconstruction results after masking an entire marker for \citet{zhu2025spatial}, \citet{schulz2024immucanpanel1}, \citet{rigamonti2024integrating}, \citet{hu2023single} and \citet{cords2023cancer}. Each row depicts different channels of the same tissue niche.}
    \label{suppfig:examples_marker_2}
\end{figure*}

\begin{figure*}[h!]
    \centering
    \includegraphics[width=\textwidth]{figures/supplement/examples_marker_3.pdf}
    \caption{Examples of image reconstruction results after masking an entire marker for \citet{moldoveanu2022spatially}, \citet{schulz2018simultaneous}, \citet{allam2022spatially} and \citet{wang2023spatial}. Each row depicts different channels of the same tissue niche.}
    \label{suppfig:examples_marker_3}
\end{figure*}

\begin{figure*}[h!]
    \centering
    \includegraphics[width=\textwidth]{figures/supplement/examples_niche_1.pdf}
    \caption{Examples of reconstruction results using niche masking, with masking ratio between 60\% and 100\%, for \cords, \damond, \danenberg, \hoch and \jackson. Each row depicts different channels of the same tissue niche.}
    \label{suppfig:examples_niche_1}
\end{figure*}

\begin{figure*}[h!]
    \centering
    \includegraphics[width=\textwidth]{figures/supplement/examples_niche_2.pdf}
    \caption{Examples of reconstruction results using niche masking, with masking ratio between 60\% and 100\%, for \citet{zhu2025spatial}, \citet{schulz2024immucanpanel1}, \citet{rigamonti2024integrating}, \citet{hu2023single} and \citet{cords2023cancer}. Each row depicts different channels of the same tissue niche.}
    \label{suppfig:examples_niche_2}
\end{figure*}

\begin{figure*}[h!]
    \centering
    \includegraphics[width=\textwidth]{figures/supplement/examples_niche_3.pdf}
    \caption{Examples of reconstruction results using niche masking, with masking ratio between 60\% and 100\%, for \citet{moldoveanu2022spatially}, \citet{schulz2018simultaneous}, \citet{allam2022spatially} and \citet{wang2023spatial}. Each row depicts different channels of the same tissue niche.}
    \label{suppfig:examples_niche_3}
\end{figure*}

\begin{figure*}[h!]
    \centering
    \includegraphics[width=\textwidth]{figures/supplement/reconstruction_losses.pdf}
    \caption{Overview of marker-wise reconstruction performance, measured by Pearson correlation, for independent (circles), marker (crosses), and niche (triangles) masking schemes. For comparison, light gray squares represent the correlation obtained by predicting the mean channel intensity of visible pixels for all masked pixels under independent masking. Dark gray diamonds indicate the correlation of each marker with the highest correlated other marker.}
    \label{suppfig:reconstruction_losses}
\end{figure*}

\begin{figure*}[h!]
    \centering
    \includegraphics[width=\textwidth]{figures/supplement/reconstruction_f1_scores.pdf}
    \caption{Overview of marker-wise reconstruction performance, measured by F1-scores of pixel-wise marker positivity, for independent (circles), marker (crosses), and niche (triangles) masking schemes. Positivity is determined by thresholding at the mean intensity of each marker. For comparison, light gray squares represent the performance obtained by predicting the mean channel intensity of visible pixels for all masked pixels under independent masking. Dark gray diamonds indicate the performance when predicting for each marker the highest correlated other marker.}
    \label{suppfig:reconstruction_f1_scores}
\end{figure*}

\begin{figure*}[h!]
    \centering
    \includegraphics[width=\textwidth]{figures/supplement/examples_zeroshot_rigamonti_known.pdf}
    \caption{Examples of zero-shot reconstruction results using independent (left), marker (middle) and niche (right) masking for markers of \citet{rigamonti2024integrating} observed during pretraining. For comparison, the same samples are also reconstructed using a model instance pretrained on \citet{rigamonti2024integrating}.}
    \label{suppfig:examples_zeroshot_known}
\end{figure*}

\begin{figure*}[h!]
    \centering
    \includegraphics[width=\textwidth]{figures/supplement/examples_zeroshot_rigamonti_new.pdf}
    \caption{Examples of zero-shot reconstruction results using independent (left), marker (middle) and niche (right) masking for markers of \citet{rigamonti2024integrating} not observed during pretraining. For comparison, the same samples are also reconstructed using a model instance pretrained on \citet{rigamonti2024integrating}.}
    \label{suppfig:examples_zeroshot_new}
\end{figure*}

\begin{figure*}[h!]
    \centering
    \includegraphics[width=\textwidth]{figures/supplement/support_cells_coarse.pdf}
    \caption{Support distribution for coarse cell types across the test sets of \cords, \citet{wang2023spatial}, \hoch, \danenberg, and \citet{rigamonti2024integrating}, illustrating the imbalance in cell type representation in cell classification tasks.}
    \label{suppfig:support_cells_coarse}
\end{figure*}

\begin{figure*}[h!]
    \centering
    \includegraphics[width=\textwidth]{figures/supplement/f1_cells_finegrained.pdf}
    \caption{F1-scores for fine-grained cell-type classification on \cords and \citet{wang2023spatial}.}
    \label{suppfig:f1_cells_finegrained.pdf}
\end{figure*}

\begin{figure*}[h!]
    \centering
    \includegraphics[width=\textwidth]{figures/supplement/support_cells_finegrained.pdf}
    \caption{Support distribution for fine-grained cell types across the test sets of \cords and \citet{wang2023spatial}, highlighting the imbalance in cell type representation in cell classification tasks.}
    \label{suppfig:support_cells_finegrained}
\end{figure*}

\begin{figure*}[h!]
    \centering
    \includegraphics[width=\textwidth]{figures/supplement/danenberg_survival_based_groups.pdf}
    \caption{\textbf{a,} Kaplan-Meier survival curves for survival-derived control risk groups on \danenberg. Risk groups are formed by balanced grouping of (i) survivors and late-censored patients, and (ii) deceased and early-censored patients. \textbf{b,} Risk ratio of each TME structure’s occurrence in the high risk control group. Lines indicate the 95\% confidence intervals.}
    \label{suppfig:danenberg_survival_based_groups}
\end{figure*}

\begin{figure*}[h!]
    \centering
    \includegraphics[width=\textwidth]{figures/supplement/support_tissue_tasks.pdf}
    \caption{Support distribution of labels for tissue-level classification tasks across the test sets of \cords, \citet{wang2023spatial} and \danenberg.}
    \label{suppfig:support_tissue_tasks}
\end{figure*}

\begin{figure*}[h!]
    \centering
    \includegraphics[width=\textwidth]{figures/supplement/attention_scores.pdf}
    \caption{Visualization of spatial and marker attention in \cords. \textbf{a, b,} Spatial attention maps derived from an extended version of \our\ incorporating an additional class token that attends to all patch summary tokens. The model was fine-tuned for cancer subtype prediction and the shown maps represent the class token’s attention scores. \textbf{c,} Marker attention scores computed across three tissue niches with distinct cell type compositions. The five markers most attended to by other markers (after averaging attention scores across these regions) are shown. Bars are color-coded according to the canonical cell type associated with each marker.}
    \label{suppfig:attention_scores}
\end{figure*}

\begin{figure*}[h!]
    \centering
    \includegraphics[width=\textwidth]{figures/supplement/examples_retrieval.pdf}
    \caption{Examples of retrieval results for tissues in \cords using the Wasserstein distance of \our's niche representation tokens as the similarity measure. Tissues are visualized using their color-coded cell type masks. Colorbars show the proportional cell type compositions.}
    \label{suppfig:retrieval_examples_cords}
\end{figure*}

\begin{figure*}[h!]
    \centering
    \includegraphics[width=\textwidth]{figures/supplement/wang_cluster_cell_type_rel_enrichment.pdf}
    \caption{Relative enrichment of cell types in selected response and non-response clusters of \citet{wang2023spatial} compared to overall proportions.}
    \label{suppfig:wang_cluster_cell_type_rel_enrichment}
\end{figure*}

\begin{figure*}[h!]
    \centering
    \includegraphics[width=\textwidth]{figures/supplement/baseline_survival_curves_meyer.pdf}
    \caption{Kaplan–Meier curves of disease-free survival on \citet{meyer2025stratification} using different baseline risk stratification systems. From left to right: (1) original risk groups from \citet{meyer2025stratification}, (2) risk scores based on tertile-transformed CD8 T cell–to–tumor ratios, (3) CD4 T cell–to–tumor ratios, and (4) B cell–to–tumor ratios. Each plot reports the p-value from a log-rank test comparing the low- and high-risk groups.}
    \label{suppfig:baseline_survival_curves_meyer}
\end{figure*}

\clearpage
\newpage
\section*{Supplementary Tables}

\begin{table}[htb]
\centering
\caption{\textbf{Overview of IMC datasets.} Only images used for the training and evaluation of \our are considered. Cell counts are indicated only for datasets with single cell annotations and masks.}
\label{table:datasets}
\begin{tabular}{l|m{5cm}cccc}
\toprule
\textbf{Dataset} & \textbf{Tissue Origin} &  \textbf{Images} & \textbf{Patients} & \textbf{Crops} & \textbf{Cells} \\
\midrule
\citet{cords2024cancer} & primary lung cancer & 1'969 & 1'048 & 50'236 & 5'880'664\\
\\
\citet{danenberg2022breast} & primary breast cancer & 688 & 688 & 11'837 & 938'561\\
\\
\citet{hoch2022multiplexed} & primary and metastatic melanoma cancer & 50 & 35 & 2'015 & 325'881\\
\\
\citet{jackson2020single} & primary breast cancer & 711 & 284 & 15'788 & 1'237'470\\
\\
\citet{damond2019map} & healthy and diabetic pancreas & 843 & 12 & 10'013 & 1'775'386\\
\\
\citet{zhu2025spatial} & primary lung cancer & 1'645 & 62 & 66'982 & -\\
\\
\citet{schulz2024immucanpanel1} & primary lung, breast, colon, kidney, head and neck cancer & 179 & 179 & 3'544 & -\\
\\
\citet{rigamonti2024integrating} & primary lung cancer & 158 & 83 & 25'586 & 659'561\\
\\
\citet{cords2023cancer} & primary breast cancer & 108 & 12 & 6'281 & -\\
\\
\citet{hu2023single} & primary lung cancer, tonsil and endometrium tissue & 51 & 12 & 4'112 & -\\
\\
\citet{moldoveanu2022spatially} & primary and metastatic melanoma cancer & 90 & 90 & 4'584 & 227'592\\
\\
\citet{schulz2018simultaneous} & primary breast cancer & 77 & 77 & 1'758 & -\\
\\
\citet{allam2022spatially} & primary lung cancer & 26 & 26 & 2'073 & -\\
\\
\citet{wang2023spatial} & primary breast cancer & 1'842 & 279 & 41'480 & 2'538'185\\
\\
\citet{meyer2025stratification} & primary breast cancer & 450 & 215 & 13'459 & 928'993\\
\\
\midrule
Total & - & 8'887 & 3'102 & 259'748 & $\geq$14'512'293\\
\bottomrule
\end{tabular}
\end{table}

\begin{table}[htb]
\centering
\caption{Nested subsets of markers from \cords, as used in the analysis of Fig.~\ref{fig:overview}e, grouped according to their presumed informativeness for general tissue characterization.}
\label{table:subsets_marker_scaling}
\begin{tabular}{l l l l}
\toprule
\textbf{Top 10 Markers} & \textbf{Top 11--20 Markers} & \textbf{Top 21--30 Markers} & \textbf{Top 31--40 Markers} \\ 
\midrule
FAP & ACTA2 & MPO & TCF7 \\
CD68 & HLA-DRA & S100A4 & MME \\
CD3E & CA9 & H3C1 & CD45RA \\
PDCD1 & MS4A1 & MCAM & CD248 \\
NT5E & IDO1 & CDH11 & LYVE1 \\
VIM & MMP9 & COL1A1 & VWF \\
FOXP3 & PDGFRB & VCAM1 & CXCL12 \\
CD8A & CD34 & PDPN & CCL21 \\
KRT14 & CD4 & MMP11 & CDH6 \\
MKI67 & FUT4 & NGFR & CAV1 \\
\bottomrule
\end{tabular}
\end{table}

\begin{table}[ht]
\centering
\caption{Grouping of the original cell phenotypes available in the metadata of \danenberg into seven high-level classes used in the cell classification task.} \label{table:mapping_celltype_danenberg}
\begin{tabular}{l l}
\toprule
\textbf{High-Level Class} & \textbf{Original Cell Phenotypes} \\ \midrule
\multirow{5}{*}{Stromal Cell} & Fibroblasts \\ 
                         & Myofibroblasts \\ 
                         & Endothelial \\ 
                         & Myofibroblasts PDPN$^{+}$ \\ 
                         & Fibroblasts FSP1$^{+}$ \\ \midrule
\multirow{5}{*}{Myeloid}     & CD15$^{+}$ \\ 
                         & Macrophages \\ 
                         & Granulocytes \\ 
                         & MHC$^{\mathrm{high}}$CD15$^{+}$ \\ \midrule
\multirow{2}{*}{Natural Killer Cell} & MHC I$^{\mathrm{high}}$CD57$^{+}$ \\ 
                         & CD57$^{+}$ \\ \midrule
\multirow{6}{*}{ER-Positive Epithelial Cell} & CK$^{+}$ CXCL12$^{+}$ \\ 
                         & ER$^{\mathrm{high}}$CXCL12$^{+}$ \\ 
                         & HER2$^{+}$ \\ 
                         & Ep Ki67$^{+}$ \\ 
                         & CK8-18$^{+}$ ER$^{\mathrm{high}}$ \\ 
                         & CK$^{\mathrm{low}}$ER$^{\mathrm{med}}$ \\ \midrule
\multirow{6}{*}{ER-Negative Epithelial Cell} & CK$^{\mathrm{med}}$ER$^{\mathrm{low}}$ \\ 
                         & CK$^{\mathrm{low}}$ER$^{\mathrm{low}}$ \\ 
                         & CK8-18$^{\mathrm{high}}$CXCL12$^{\mathrm{high}}$ \\ 
                         & Basal cells\\ 
                         & CK8-18$^{\mathrm{high}}$ER$^{\mathrm{low}}$ \\ 
                         & Ep CD57$^{+}$ \\ \midrule
\multirow{3}{*}{T Cell}     & CD8$^{+}$ T cell \\ 
                         & CD4$^{+}$ T cell \\ 
                         & T$_{Reg}$ and T$_{Ex}$ cells \\ \midrule
\multirow{2}{*}{B Cell}     & B cells \\
                         & CD38$^{+}$ lymphocytes \\ \midrule
 \multirow{1}{*}{Antigen-Presenting Cell} & MHC I$^{\mathrm{high}}$ \& II$^{\mathrm{high}}$ \\ \bottomrule
\end{tabular}
\end{table}

\begin{table}[ht]
\centering
\caption{Grouping of the original cell phenotypes from the \hoch metadata into five high-level classes used in the cell classification task.} \label{table:mapping_celltype_hoch}
\begin{tabular}{l l}
\toprule
\textbf{High-Level Class} & \textbf{Original Cell Phenotypes} \\ \midrule
Tumor & Tumor, HLA-DR \\ \midrule
T Cell & CD8\textsuperscript{+} T cells, CD8\textsuperscript{-} T cells \\ \midrule
Macrophage              & Macrophage \\\midrule
Lymphocyte & CD38, Neutrophil \\ \midrule
Stroma & Stroma, Vasculature \\ \midrule
Other & Other \\
\bottomrule 
\end{tabular}
\end{table}

\begin{table}[ht]
\centering
\caption{Grouping of the original cell phenotypes available in the metadata of \citet{wang2023spatial} into six high-level classes and 19 low-level classes used in the cell classification task.} \label{table:mapping_celltype_wang}
\begin{tabular}{l l l}
\toprule
\textbf{High-Level Class} & \textbf{Low-Level Class} & \textbf{Original Cell Phenotypes} \\ 
\midrule
\multirow{3}{*}{Stroma} & \multirow{3}{*}{Fibroblasts} & Fibroblasts \\ 
                       &                              & PDPN$^{+}$Stromal \\ 
                       &                              & Myofibroblasts \\ 
\midrule
\multirow{12}{*}{Immune} & \multirow{3}{*}{Dendritic / APC} & DCs \\ 
                        &                                  & PD-L1$^{+}$APCs \\ 
                        &                                  & PD-L1$^{+}$IDO$^{+}$APCs \\ 
                        & Macrophages                      & M2 Mac \\ 
                        & \multirow{2}{*}{B / Plasma cells} & CD79a$^{+}$Plasma \\ 
                        &                                  & CD20$^{+}$B \\ 
                        & \multirow{2}{*}{NK cells}        & CD56$^{+}$NK \\ 
                        &                                  & CD56$^{+}$NE \\ 
                        & \multirow{2}{*}{Neutrophils}     & CD15$^{+}$ \\ 
                        &                                  & Neutrophils \\ 
\midrule
\multirow{12}{*}{Tumor} & \multirow{2}{*}{Hypoxic Tumor}    & CA9$^{+}$ \\ 
                        &                                  & CA9$^{+}$Hypoxia \\ 
                        & \multirow{6}{*}{Tumor cells}     & CK8/18$^{\text{med}}$ \\ 
                        &                                  & panCK$^{\text{med}}$ \\ 
                        &                                  & CK$^{\text{hi}}$GATA3$^{+}$ \\ 
                        &                                  & CK$^{\text{lo}}$GATA3$^{+}$ \\ 
                        &                                  & Basal \\ 
                        &                                  & AR$^{+}$LAR \\ 
                        & DNA-damage cells                 & pH2AX$^{+}$DSB \\ 
                        & EMT-like Tumor                   & Vimentin$^{+}$EMT \\ 
                        & Apoptotic cells                  & Apoptosis \\ 
\midrule
\multirow{9}{*}{T Cell} & \multirow{2}{*}{CD4 T cells}      & CD4$^{+}$TCF1$^{+}$T \\ 
                        &                                  & CD4$^{+}$PD1$^{+}$T \\ 
                        & \multirow{2}{*}{Regulatory T cells} & Treg \\ 
                        &                                     & Helios$^{+}$ \\ 
                        & \multirow{4}{*}{CD8 T cells}     & CD8$^{+}$T \\ 
                        &                                  & CD8$^{+}$TCF1$^{+}$T \\ 
                        &                                  & CD8$^{+}$PD1$^{+}$T$_{\text{Ex}}$ \\ 
                        &                                  & CD8$^{+}$GZMB$^{+}$T \\ 
                        & Stem-like T cells                & TCF1$^{+}$ \\ 
\midrule
Vessel                  & Endothelial                      & Endothelial \\ 
\midrule
\multirow{3}{*}{Other}  & PD-L1$^{+}$GZMB$^{+}$ cells       & PD-L1$^{+}$GZMB$^{+}$ \\ 
                        & PD-L1$^{+}$IDO$^{+}$ cells        & PD-L1$^{+}$IDO$^{+}$ \\ 
                        & MHC-I\&II$^+$ cells                 & MHC I\&II$^{\text{hi}}$ \\ 
\bottomrule
\end{tabular}
\end{table}

\begin{table}[ht]
\centering
\caption{Grouping of the original cell phenotypes available in the metadata of \citet{rigamonti2024integrating} into six high-level classes used in the cell label transfer experiment.} \label{table:mapping_celltype_rigamonti}
\begin{tabular}{l l}
\toprule
\textbf{High-Level Class} & \textbf{Original Cell Phenotypes} \\ \midrule
Tumor & Tumor cells \\ \midrule
T cell & CD8+, CD4+, Treg \\ \midrule
Immune & B cells, Mf, Ki67+ \\\midrule
Fibroblast & aSMA+, Vim \\ \midrule
Vessel & Vessels \\ \midrule
Other & Other \\
\bottomrule 
\end{tabular}
\end{table}

%% file: boxes.tex
\newcounter{theo}[section] 
\renewcommand{\thetheo}{\arabic{theo}}

%% file: content/abstract.tex
\begin{abstract}
Spatial proteomics technologies have transformed our understanding of complex tissue architecture in cancer but present unique challenges for computational analysis. Each study uses a different marker panel and protocol, and most methods are tailored to single cohorts, which limits knowledge transfer and robust biomarker discovery. Here we present Virtual Tissues (VirTues), a general-purpose foundation model for spatial proteomics that learns marker-aware, multi-scale representations of proteins, cells, niches and tissues directly from multiplex imaging data. From a single pretrained backbone, VirTues supports marker reconstruction, cell typing and niche annotation, spatial biomarker discovery, and patient stratification, including zero-shot annotation across heterogeneous panels and datasets. In triple-negative breast cancer, VirTues-derived biomarkers predict anti-PD-L1 chemo-immunotherapy response and stratify disease-free survival in an independent cohort, outperforming state-of-the-art biomarkers derived from the same datasets and current clinical stratification schemes. 
\end{abstract}

%% file: content/introduction.tex
\section*{Introduction}\addcontentsline{toc}{section}{Introduction}

Tissues, particularly in cancer, display pronounced heterogeneity across patients, disease stages and even within individual tumors, evident in diverse cell phenotypes, states and spatial organization \citep{de2024multiplex}. Accounting for this heterogeneity is critical: tumor development and response to therapy depend not just on cancer cells, but on their complex interactions with their surrounding environment. Different cell phenotypes may act as promoters or suppressors of tumor development and progression, depending on the biological context \citep{hanahan2011hallmarks, hanahan2022hallmarks}, with their spatial co-occurrence patterns predicting immunotherapy response \citep{wang2023spatial, phillips2021immune}, disease relapse \citep{radtke2024multi} and survival \citep{sorin2023single}. Understanding the spatial organization, composition and function of the tumor microenvironment (TME) has thus emerged as an important element for advancing cancer treatment.

\begin{figure*}[ht!]
    \centering
    \includegraphics[width=\textwidth]{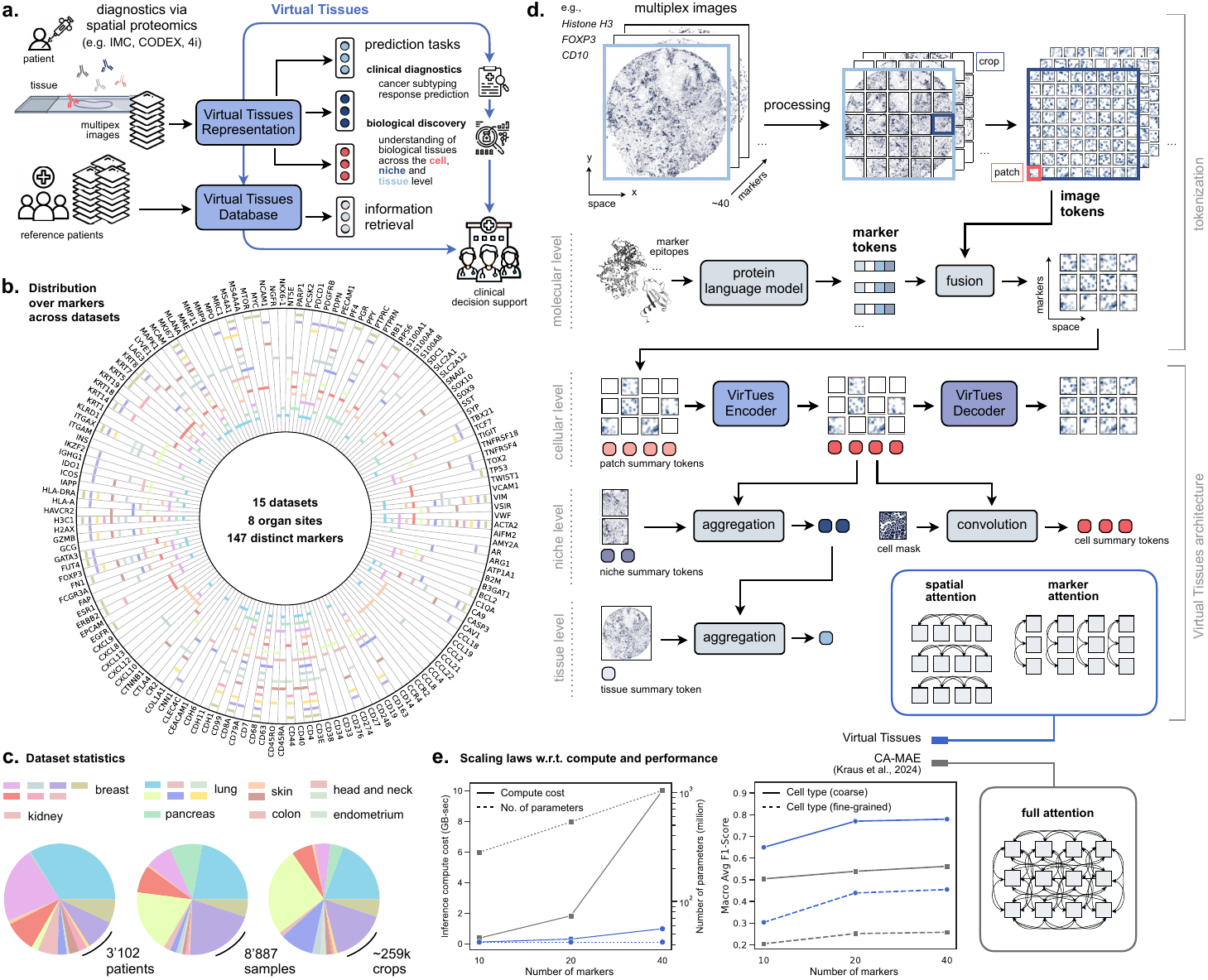}
    \caption{\textbf{Overview of the Virtual Tissues platform}. 
    \textbf{a,} Flow chart depicting \our\ capabilities. \our\ converts highly multiplexed images of tissue to virtual tissue representations useful for clinical and biological investigations at cell, niche and sample level including the retrieval of similar tissue samples for clinical decision support. 
    \textbf{b,} \our\ is trained and evaluated on 15 IMC datasets with a focus on tumors and their micro-environments originating from 8 different organ sites, measuring 147 distinct markers in total. The polar plot depicts used marker panels per dataset. A legend of the dataset color codes is provided in Suppl. Fig. \ref{suppfig:legend_datasets}.
    \textbf{c,} Origins and sizes of datasets in terms of patients, tissue samples, and $256\times256$ image crops.
    \textbf{d,} Multiplexed images are processed crop-wise into 3D grids of image tokens, representing patches of each marker at each position. Marker tokens, derived from a protein language model, are fused with the respective image tokens using a linear projection and addition. \our\ is a novel vision transformer architecture trained with a masked autoencoding objective. Input tokens are concatenated with patch summary tokens, which are initialized with learnable weights. During inference, \our' encoder processes this set of tokens. The encoded patch summaries are subsequently convolved with the cell segmentation mask into cell summary tokens or aggregated to niche and tissue summary tokens. For training, a random subset of tokens is independently selected and masked for each channel. \our' decoder predicts channel-wise reconstructions receiving as input the encoded, non-masked tokens from the target channel along with all patch summary tokens. \our\ encoder uses sparse attention mechanisms restricting direct token interactions to either positions (marker attention) or channels (spatial attention).
    \textbf{e,} Comparison of computational cost (left) and prediction performance (right) between channel-agnostic masked auto-encoder (\camae) and \our\ as a function of the number of utilized markers.}
    \label{fig:overview}
\end{figure*}

Capturing complex tissue structure requires advanced molecular imaging techniques that go beyond traditional methods \citep{lin2023high}. The emergence of multiplexed imaging technologies---including conventional immunohistochemistry, multiplexed immunofluorescence and imaging mass cytometry---has enabled the simultaneous measurement of dozens to hundreds of proteins in intact tissue sections at subcellular resolution. These spatial proteomics assays have revealed intricate patterns of tumor--immune interactions, spatial niches linked to prognosis and mechanisms of resistance to therapy across multiple cancer types, and they are becoming central to basic cancer biology, biomarker discovery and the design of rational combination therapies \citep{danenberg2022breast} (Fig.~\ref{fig:overview}a).

However, the same flexibility that makes multiplexed imaging powerful also creates major analytical challenges. Each study typically relies on a customized panel of antibodies, staining protocol and imaging platform optimized for a specific question or tissue. The resulting datasets differ in the number and identity of markers (Fig.~\ref{fig:overview}b), dynamic range, and noise characteristics. Current computational workflows are often designed for a single cohort or panel, with dedicated pipelines for segmentation, normalization, clustering and analysis of cell--cell interactions. As a consequence, it remains difficult to transfer knowledge between cohorts, cancer types or platforms, to reuse well-annotated reference datasets, or to establish spatial biomarkers that generalize beyond the study in which 
they were discovered \citep{tsimberidou2023molecular,rigamonti2024integrating}. This problem is particularly acute when biomarkers are intended to inform treatment decisions or stratify patients in clinical trials.

In parallel, recent years have seen the emergence of foundation models in natural language processing\citep{openai2023gpt4, brown2020language, touvron2023llama, gemini2023}, computer vision\citep{dosovitskiy2021an, he2022masked, radford2021learning, bachmann2022multimae}, and histopathology\citep{chen2024towards, xu2024whole, wang2024pathology}, where a single model trained at scale provides reusable representations for many downstream tasks. In oncology, such models have started to support diagnosis, grading and outcome prediction from routine pathology images and radiology scans. Yet there is no general-purpose foundation model that operates directly on high-plex spatial proteomics data across heterogeneous marker panels while remaining biologically interpretable. Existing encoders for multiplexed imaging typically require fixed marker panels \citep{sorin2023single, gupta2024subcell, pfaendler2023self}, exhibit limited scalability to highly multiplexed datasets \citep{kraus2024masked, kenyon2024vitally}, or are evaluated on narrow tasks rather than on cross-cohort biomarker discovery and clinical stratification \citep{bao2023channel, sorin2023single, wang2024generalized, shaban2025foundation, farndale2025self}. Moreover, most approaches do not explicitly leverage molecular priors about protein structure and function when learning spatial representations.

Here we introduce Virtual Tissues, a marker-aware, multi-scale foundation model for spatial proteomics that addresses these limitations. VirTues is trained on a large collection of IMC and related datasets spanning multiple cancer types and centers (Fig.~\ref{fig:overview}c). At its core, \our\ combines protein-level embeddings derived from protein language models with a factorized Transformer architecture (Fig.~\ref{fig:overview}d) that through its novel attention mechanism scales to high-dimensional multiplex data while maintaining interpretability. The model produces unified representations at the level of molecules, cells, niches and whole tissues, and can efficiently process arbitrary, study-specific marker panels at inference time (Fig.~\ref{fig:overview}e). From a single pretrained backbone, VirTues supports marker reconstruction, zero-shot cell typing and niche annotation, tissue retrieval based on spatial phenotypes and, critically, the discovery of spatial biomarkers that transfer across cohorts.

We demonstrate that VirTues can integrate heterogeneous spatial proteomics datasets into a shared representational space, enabling cross-study analysis and the integration of novel markers. In triple-negative breast cancer\citep{wang2023spatial}, we show that FM-derived spatial signatures learned by VirTues predict anti--PD-L1 chemo-immunotherapy response and stratify disease-free survival in an independent cohort\citep{meyer2025stratification}, outperforming previously reported spatial biomarkers\citep{wang2023spatial} on the same datasets as well as current clinical stratification schemes. Together, these results position VirTues as a reusable computational layer for spatial proteomics and a blueprint for foundation models that couple detailed tissue understanding with clinically relevant biomarker discovery.

%% file: content/results.tex
\section*{Results}\addcontentsline{toc}{section}{Results}

\subsection*{A novel vision transformer to construct AI-powered Virtual Tissues}

\looseness -1 A pivotal characteristic of foundation models is their ability to leverage larger and more diverse training datasets in order to achieve superior performance across diverse downstream tasks. However, current vision models for biological imaging \citep{kraus2024masked, kenyon2024vitally, gupta2024subcell, bao2023channel, doron2023unbiased}, including Vision Transformers\citep{dosovitskiy2021an} (ViTs), face several limitations when applied to multiplexed biomedical data (Fig.~\ref{fig:overview}e) constraining their scalability to large and heterogeneous collections of multiplexed datasets. 
First, the current architectures' computational complexity scales quadratically with spatial dimensions and channel number, making them impractical for high-dimensional spatial data. Second, their token-based representations typically treat all channels equally, failing to capture important marker-specific information and to ignore potentially redundant biological information from different markers. Third, they lack explicit mechanisms for integrating datasets with differing marker panels (e.g., antibody panels) as well as the ability to incorporate prior knowledge about characteristics of the probed molecules (e.g., protein interactions).

We specifically design \our\ to overcome the identified challenges and introduce a purpose-built ViT model for multiplexed imaging of proteins and mRNAs. By disentangling the transformer's attention mechanism \citep{vaswani2017attention,bertasius2021space} into marker and spatial attention components, \our\ can be trained on images with dozens of channels (Fig.~\ref{fig:overview}e), and can separately learn both the spatial arrangement and cellular composition that shape a tissue's molecular profile and the interrelations between proteins and RNA markers. Another critical innovation in \our\ is the introduction of a new tokenization scheme for spatial biology: by combining protein language model embeddings \citep{lin2023evolutionary} with spatially-patched channel information and learnable patch summary tokens, we enable flexible processing of variable marker combinations while incorporating biological meaning and subcellular spatial distribution of markers (Fig.~\ref{fig:overview}d, Methods). This tokenization and multi-scale design \citep{bunne2024build, rosen2023universal} thus allows to incorporate markers not seen during training and enables the integration of heterogeneous datasets measured with different marker combinations (Fig.~\ref{fig:overview}b). Here, we train and evaluate \our\ on a collection of 15 IMC datasets, mapping a total 147 distinct markers (proteins, protein modifications, and mRNAs). These datasets sampled from 8 different organ sites show substantial variation in patient numbers (3'102 total patients), samples (8'887 total tissues), measured tissue area (over 259'000 $256\times256$ crops) and cell counts (over 14.5 million segmented cells across nine datasets) (Fig.~\ref{fig:overview}c and Suppl. Table~\ref{table:datasets}).

Building such multiplex-aware tokenization schemes and architectures is critical: modality-agnostic designs used previously neither scale nor capitalize on additional markers, whereas \our\ improves consistently with marker depth, especially for the first 20 markers (Fig.~\ref{fig:overview}e).

Based on the masked autoencoder (MAE) framework\citep{he2022masked}, \our' architecture consists of an Encoder and Decoder trained jointly in an unsupervised manner to reconstruct partially masked marker-space tensors (Fig.~\ref{fig:masking}, Methods). By learning to recover missing information, \our\ captures rich patterns in multiplex imaging data across subcellular, cellular and multicellular scales. 
Leveraging this learned representation, \our\ generates patch, cell, niche and tissue summary tokens, providing a hierarchical and multi-scale tissue representation. These tokens enable downstream predictions across cellular, niche and tissue levels, supporting a variety of tasks from cell annotation, marker inpainting, to clinical predictions and biological discovery (Figs.~\ref{fig:cell_annotation}, \ref{fig:clinical} and \ref{fig:discovery}).

\subsection*{\our\ learns tissue architectures and relations between markers}

\begin{figure*}[ht!]
    \centering
    \includegraphics[width=\textwidth]{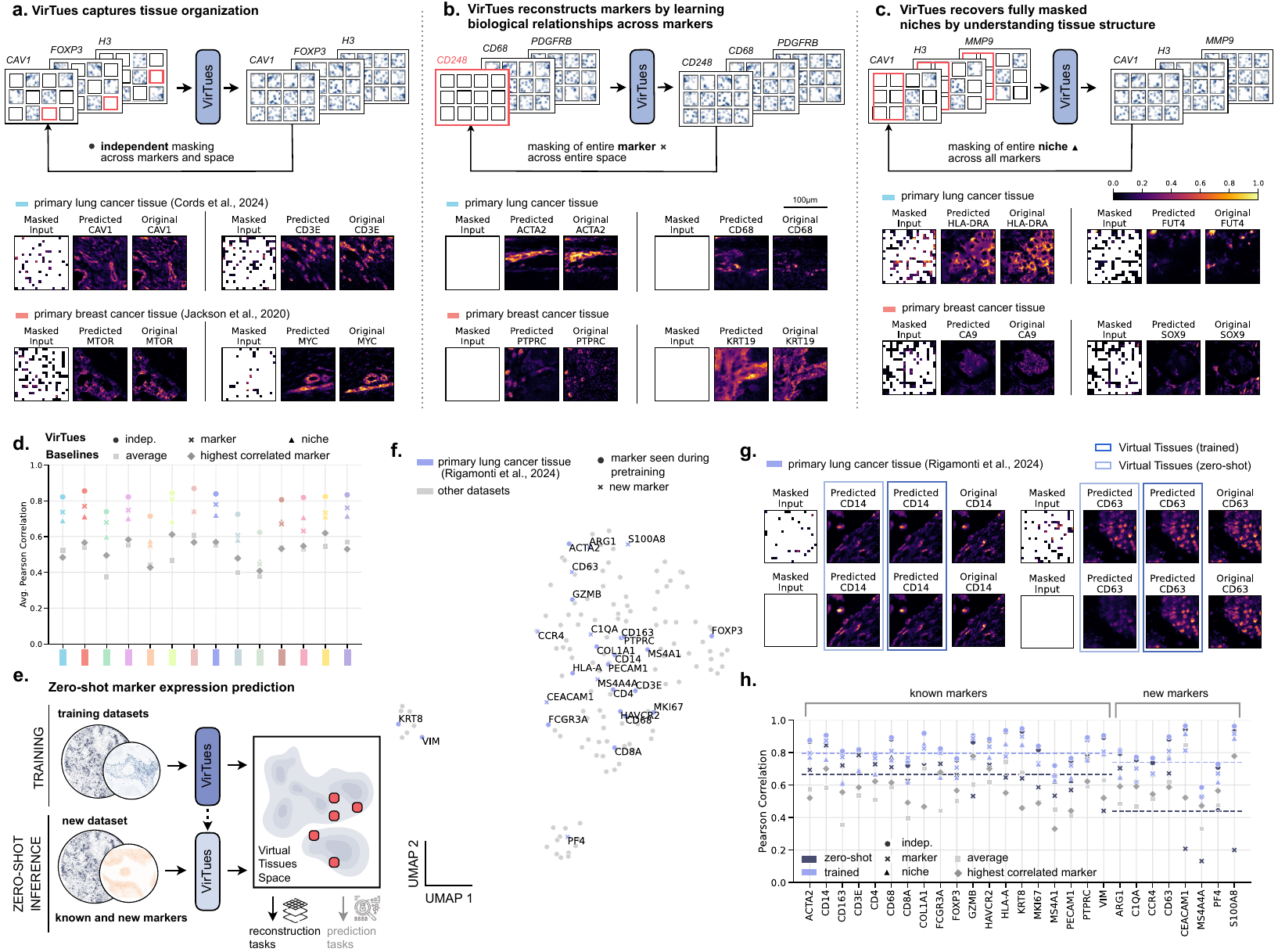}
    \caption{\textbf{\our\ learns tissue architecture and marker relationships.} 
    \textbf{a-c,} Illustrations of masking strategies and reconstruction examples. Visualized images are scaled to the interval $[0,1]$ by reversing the standardization and log-transformation of the preprocessing and dividing by the 99th percentile.
    \textbf{a,} Independent masking. Tokens of each channel are masked independently, with a channel-wise random masking ratio ranging from 60\% to 100\%.
    \textbf{b,} Marker masking. One marker is chosen and all tokens of its channel are masked while all other channels remain unmasked. In the examples, each row depicts the inpainting of different channels from the same tissue sample.
    \textbf{c,} Niche masking. A subset of spatial positions is chosen and all tokens across markers are masked at these positions. Each row depicts the niche reconstructions of different channels of the same tissue sample.
    \textbf{d,} Dataset-wise reconstruction performance quantified by Pearson correlation and averaged across markers for independent (round), marker (cross) and niche (triangle) masking. For comparison, light gray squares represent the correlation obtained by predicting the mean channel intensity of visible pixels for all masked pixels under independent masking. Dark gray diamonds indicate the correlation achieved by predicting for each marker the highest correlated other marker. Contrary to the training objective, the reconstruction performance is assessed only on masked tokens. A legend of the dataset color codes is provided in Supp. Fig. \ref{suppfig:legend_datasets}.
    \textbf{e,} \our' architecture enables the generation of virtual tissue representations for multiplexed images from new 
    datasets, including those with markers not observed during pretraining. These representations support a variety of reconstruction and prediction tasks.
    \textbf{f,} UMAP embeddings of all marker tokens derived from ESM-2. Markers, included in the panel of \citet{rigamonti2024integrating} are displayed in blue, all others in light-gray. Markers seen during pretraining, when training without \citet{rigamonti2024integrating}, are depicted as round, while new markers unique to \citet{rigamonti2024integrating} are shown as crosses. 
    \textbf{g,} Examples of zero-shot reconstructions for markers CD14 and CD63 from \citet{rigamonti2024integrating}, where CD14 was observed and CD63 unobserved during pretraining. For comparison, reconstructions of the same samples from a model additionally pretrained on \citet{rigamonti2024integrating} are shown.
    \textbf{h,} Comparison of marker-wise reconstruction performance in the zero-shot and non-zero-shot setting on \citet{rigamonti2024integrating}. Markers are grouped into those seen during pretraining (left) and unseen ones (right). Dashed lines indicate the averages across markers using marker masking.
    }
    \label{fig:masking}
\end{figure*}

To evaluate \our' understanding of tissue architecture, we challenged its ability to reconstruct tissues masked under three masking strategies of increasing difficulty: (1) \textit{independent masking}, where independently for each marker patches are masked at random spatial locations (Fig.~\ref{fig:masking}a), (2) \textit{marker masking}, where all patches of a single marker are masked (Fig.~\ref{fig:masking}b) and (3) \textit{niche masking}, where patches across all markers are masked at randomly selected spatial locations (Fig.~\ref{fig:masking}c).

In the independent masking experiments, \our\ successfully reconstructs masked regions across markers, e.g., CAV1 and CD3E in Fig.~\ref{fig:masking}a in lung cancer tissue \citep{cords2024cancer} as well as MTOR and MYC in breast cancer samples \citep{jackson2020single}, preserving both spatial distribution and intensity patterns (see Suppl. Figs.~\ref{suppfig:examples_independent_1}, \ref{suppfig:examples_independent_2}, \ref{suppfig:examples_independent_3} for results on further channels of all datasets). This demonstrates the model's ability to leverage contextual information across both spatial and marker dimensions.

In the more difficult marker masking experiments (Fig.~\ref{fig:masking}b), where individual channels are masked entirely, \our\ accurately recovers the expression patterns of masked markers, e.g., ACTA2 and CD68 in lung cancer \citep{cords2024cancer} as well as PTPRC and KRT19 in breast cancer tissue \citep{jackson2020single} (see Suppl. Figs.~\ref{suppfig:examples_marker_1}, \ref{suppfig:examples_marker_2}, \ref{suppfig:examples_marker_3} for results on further channels of all datasets), indicating its ability to learn biological relationships between different markers. Since with marker masking the \our\ Decoder receives only the encoded patch summary tokens as input, its ability to reconstruct a missing channel validates that these tokens capture a robust representation of the molecular tissue architecture.

Under niche masking, (Fig.~\ref{fig:masking}c), where entire tissue regions are occluded across all markers, \our\ successfully reconstructs complex tissue architectures, recovering marker patterns, e.g., HLA-DRA and FUT4 in lung cancer \citep{cords2024cancer} as well as CA9 and SOX9 in breast cancer tissue \citep{jackson2020single} (see Suppl. Figs.~\ref{suppfig:examples_niche_1}, \ref{suppfig:examples_niche_2}, \ref{suppfig:examples_niche_3} for results on further channels of all datasets). The ability to recover spatially coherent macro structures demonstrates that the model captures global tissue architecture rather than relying solely on local pixel information.

We further quantitatively confirm our observations: reconstructions using \our\ achieved high Pearson correlations (average $r = 0.723 \pm 0.157$) across markers, datasets, and masking strategies (Fig.~\ref{fig:masking}d; see Suppl. Figs. ~\ref{suppfig:reconstruction_losses} and \ref{suppfig:reconstruction_f1_scores} for marker-wise performance). Across all strategies, \our\ outperforms baselines such as reconstructions based on mean channel intensity or using the most correlated marker as a predictor, showcasing its ability to integrate contextual signals and complex marker interactions for accurate tissue reconstruction.

Finally, we assessed zero-shot generalization for tissue reconstruction on an independent lung cancer dataset\citep{rigamonti2024integrating} withheld entirely during a separate training run (Fig.~\ref{fig:masking}e). This dataset included \textit{both} markers present in the training cohorts and novel markers not observed during training (Fig.~\ref{fig:masking}f). In this setting, \our\ generated visually accurate reconstructions across previously seen and never seen before markers (e.g., CD14 and CD63, Fig.~\ref{fig:masking}g; see Suppl. Figs.~\ref{suppfig:examples_zeroshot_known} and ~\ref{suppfig:examples_zeroshot_new} for additional examples). Marker-wise quantitative analysis revealed substantial variation in reconstruction performance across both markers and masking strategies (Fig.~\ref{fig:masking}h). 
For known markers (previously seen in training cohorts), zero-shot reconstructions achieved high correlations (average $r=0.667$) despite a reduction in performance compared to a model evaluated in-domain (average $r=0.797$). Interestingly, this performance drop occurred only when entire markers were masked, but not under independent or niche masking (where performance differed only by $\Delta r=0.016$ resp. $\Delta r=-0.002$).
In the challenging setting of reconstructing \textit{new} markers present only in the zero-shot dataset, the average performance drop was more striking, again, however, mostly when masking entire markers. Moreover, zero-shot reconstruction performance under independent masking was higher than under niche masking, suggesting that even for novel markers, \our\ leverages not only local spatial patterns but also inter-marker relationships.

\subsection*{\our\ accurately annotates tissues at the single-cell scale}

\our\ enables comprehensive analysis of tissues across biological scales through a unified framework of patch, cell, niche, and tissue summary tokens (Fig.~\ref{fig:overview}d). This framework, for example, allows to annotate cell types at the single-cell level, a critical step for characterizing cellular composition, mapping how distinct cell identities contribute to the tumor microenvironment (TME), and ultimately informing therapeutic decision-making\citep{wang2023spatial,loi2020relationship,nederlof2024neoadjuvant}.

\begin{figure*}[t]
    \centering
    \includegraphics[width=\textwidth]{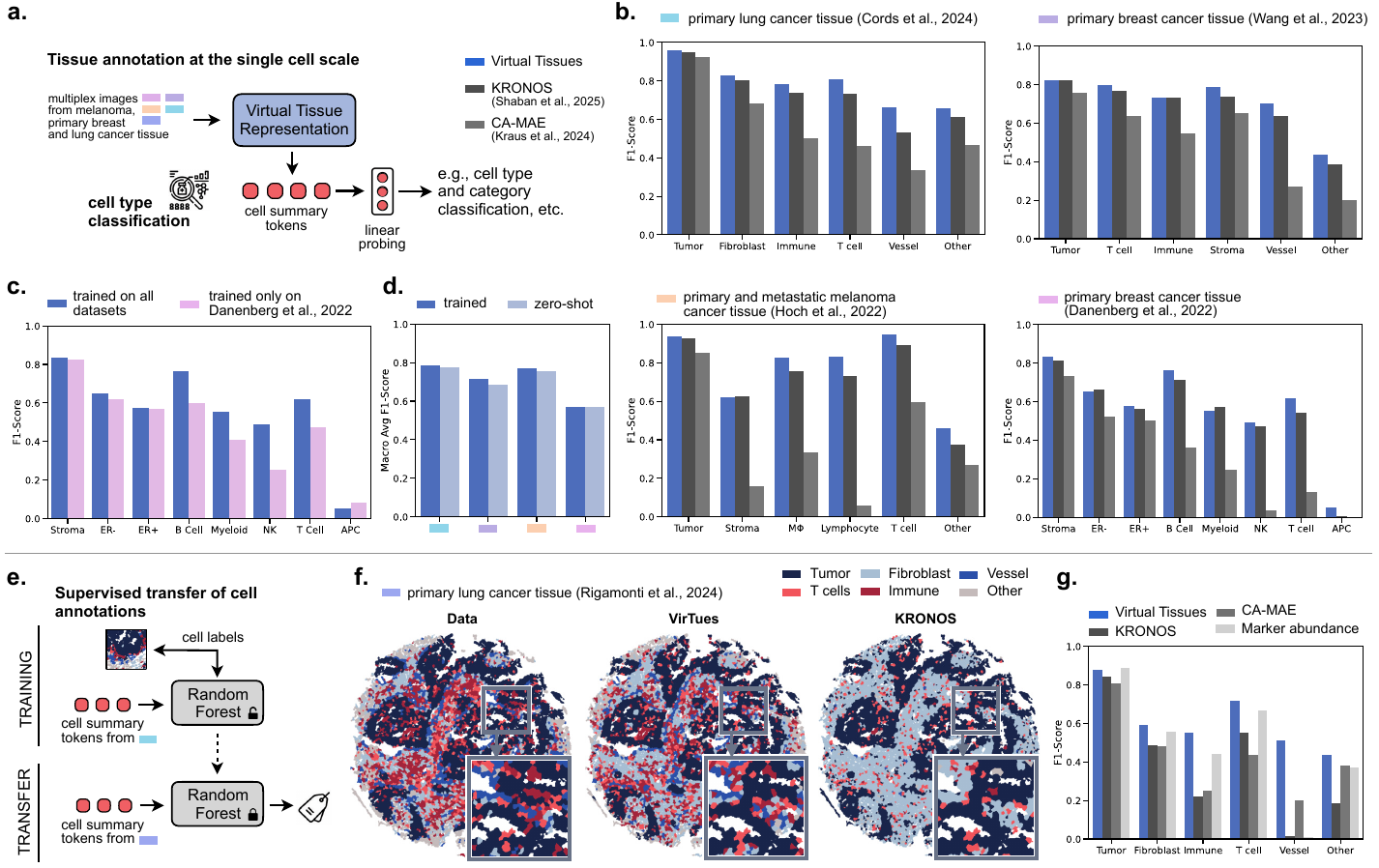}
    \caption{\textbf{Evaluation of \our' cell-level representations.}
    \textbf{a,} Cell types are classified with logistic regression using individual cell summary tokens.
    \textbf{b,} F1-scores for cell type classification for \our, \kronos and \camae on \cords, \citet{wang2023spatial}, \hoch and \danenberg. 
    \textbf{c,} Comparison of F1-scores for cell type classifications between \our\ trained an all pretraining datasets and \our\ trained only on \danenberg.
    \textbf{d,} Comparison of cell type classification performance in the zero-shot and the non-zero-shot setting measured by macro-averaged F1-score. Results are shows from left to right for \cords (coarse cell types), \citet{wang2023spatial} (coarse cell types), \hoch and \danenberg.
    \textbf{e,} Virtual Tissues facilitates the supervised transfer of annotations between datasets. We train a Random Forest classifier on the cell summary tokens of \cords to predict cell types using available labels. Subsequently, we employ the trained classifier to annotate cell summary tokens of \citet{rigamonti2024integrating} computed using the shared markers.
    \textbf{f,} Examples of a cell type masks generated using transferred labels. We compare to the transfer results using \kronos' cell-level representations (left) and ground truth (right).
    \textbf{g,} F1-scores for the transfer of cell types for \our, \kronos and \camae and mean abundances of the shared markers.
    }
    \label{fig:cell_annotation}
\end{figure*}

We evaluated \our' performance in annotating cell types. For this, we employed logistic regression, also called linear probing, on \our' cell summary tokens (Fig.~\ref{fig:cell_annotation}a). This allows assessing the discriminative performance and the representation quality of a foundation model encoder \citep{balestriero2023cookbook}. We compared against two recent state-of-the-art vision encoder models developed for microscopy images, \kronos and \camae. 
\our\ achieved strong classification performance across several lung, breast and melanoma cancer tissue datasets (Fig.~\ref{fig:cell_annotation}b).
Notably, \our\ consistently outperformed both baselines in macro-averaged F1-score, achieving average relative improvements of +6.31\% over \kronos\ and +65.79\% over \camae, while maintaining strong performance on underrepresented classes such as vessel and natural killer cells (Suppl. Fig.~\ref{suppfig:support_cells_coarse}). For instance, vessel ($<1.6\%$) and T cells ($<6.3\%$) in the \cords dataset, and  natural killer ($<1.8\%$) and B cells ($<4.9\%$) in \danenberg were exquisitely rare.
Additional classification results and support distributions for more fine-grained and highly imbalanced phenotype classes on \cords\ and \citet{wang2023spatial} are shown in Suppl. Figs.~\ref{suppfig:f1_cells_finegrained.pdf} and \ref{suppfig:support_cells_finegrained}, with \our\ maintaining consistent performance. \\

A critical property of foundation models is their ability to improve as the size and diversity of their training data increase. 
To demonstrate that \our\ exhibits this behavior, we evaluate cell type classification in \danenberg by comparing a model trained on all available datasets with one trained solely on \danenberg (Fig.~\ref{fig:cell_annotation}c).
Based on F1-scores, \our\ trained on the full corpus outperformed the single-dataset trained instance across nearly all classes. We observed the largest gains for rare immune populations such as B cells ($+27.9\%$), myeloids ($+35.2\%$), natural killer cells ($+95.6\%$) and T cells ($+30.4\%$). These results demonstrate effective cross-dataset synergies: \our\ leverages representations learned from diverse, independent cohorts to surpass a model trained only on \danenberg. \\

We further tested generalization in a zero-shot setting by training instances \our\ while entirely withholding the cohorts used for evaluation. Across all four benchmarking datasets (\cords, \citet{wang2023spatial}, \hoch and \danenberg), zero-shot classification performance remained nearly unchanged relative to in-domain training (macro-averaged F1 differences $\leq$ 0.03; Fig.~\ref{fig:cell_annotation}d). \\

The previous cell classification experiments relied on the availability of training labels within each evaluated dataset. However, \our' ability to map diverse datasets within a shared representation space also enables \textit{cross-cohort label transfer}, an especially valuable capability for annotating newly generated datasets. To test this, we trained a random forest classifier using cell type labels from \cords and applied it to \citet{rigamonti2024integrating} (Fig.~\ref{fig:cell_annotation}e). The resulting predictions overlap with ground truth annotations, as depicted in Fig.~\ref{fig:cell_annotation}f. Quantitatively, the transferred labels achieved a macro-averaged F1-score of 0.615, outperforming those obtained by \kronos (0.383) and joint mean marker intensities (0.490) (Fig.~\ref{fig:cell_annotation}g). This improvement was largely driven by a better classification of rare cell types, including immune cells, T cells, and most notably, vascular cells, reinforcing our previous findings that \our\ integrates rare but relevant information from its training corpus to solve diverse tasks salient to biomedical research.

\subsection*{\our\ representations identify TME-associated risk groups in breast cancer}
\begin{figure*}[ht!]
    \centering
    \includegraphics[width=\textwidth]{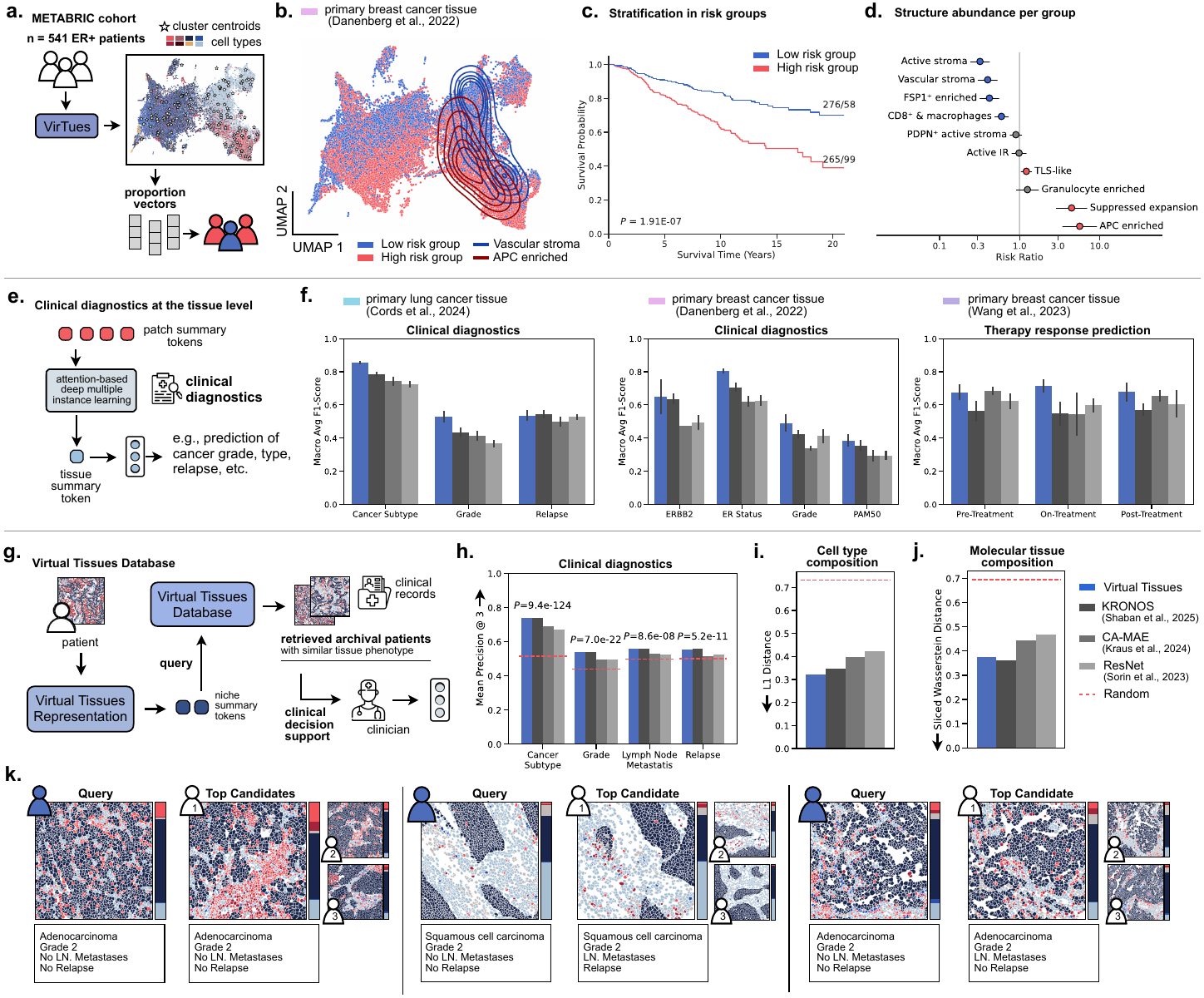}
    \caption{
    \textbf{Clinical applications of \our: risk stratification, diagnostic predictions and tissue retrieval by similarity.}
    \textbf{a-c,} \our\ enables TME-based stratification of patients into risk groups.
    \textbf{a,} Stratification protocol. Cell-level representations of ER-positive patients from the METABRIC cohort \citep{curtis2012genomic, danenberg2022breast} are computed and clustered using k-means. For each patient, a cluster proportion vector is calculated. These vectors are then clustered again via k-means to define patient groups, which are subsequently assigned risk levels. 
    \textbf{b,} UMAP embeddings of cell summary tokens, colored by the assigned patient risk level. Overlaid KDE plots show the distributions of cells from two TME structures, (1) vascular stroma and (2) APC-enriched, as previously reported by \citet{danenberg2022breast} to be associated with decreased and increased hazard ratios, respectively. 
    \textbf{c,} Kaplan-Meier survival curves of identified high-risk and low-risk groups. We report the p-value of a log-rank test comparing the two curves.
    \textbf{d,} Risk ratio of each TME structure’s occurrence in the high risk group. Lines indicate the 95\% confidence intervals.
    \textbf{e,} We predict clinical patient features from the patch summary tokens of the entire tissue using ABMIL.
    \textbf{f,} Macro-averaged F1-scores for tissue-level prediction tasks on \cords, \danenberg and \citet{wang2023spatial}. For the response prediction on \citet{wang2023spatial}, we test performance separately for biopsy samples collected before, during or after chemo- and immunotherapy. 
    \textbf{g,} \our enables data-driven clinical decision support by retrieving similar patient cases from a database of tissue representations based on \our\ niche summary tokens and an optimal transport-based retrieval system.
    \textbf{h-j}, Comparison of retrieval statistics evaluated on \cords using the niche representations of \our, \kronos, \camae\ and \resnet. The red dotted lines indicate the scores achieved by uniformly random retrieval for reference.
    \textbf{h,} Mean precision of the top-3 results for the retrieval of four clinical labels: cancer subtype, grade, presence of lymph node metastasis and relapse. P-values above the bars are computed based on a McNemar test for each clinical label and indicate the number of hits achieved by \our\ compared to a random retrieval.
    \textbf{i,} Average cell type composition similarity between query and closest match quantified by the L1 distance between the cell type proportion vectors. 
    \textbf{j,} Average molecular composition similarity between query and closest match measured by the sliced Wasserstein distance between the pixel-sized marker intensity vectors.
    \textbf{k,} Exemplary retrieval results for tissues in \cords. Each column shows the query tissue followed by the three closest matches. Tissues are depicted using their color-coded cell type masks. Colorbars next to the tissues indicate their proportional cell type compositions.
    }
    \label{fig:clinical}
\end{figure*}

A central aim of spatial biology is the study of the TME and the discovery of cellular and multi-cellular features that stratify patients by clinical risk, inform prognosis, and support therapeutic decision-making \citep{de2024multiplex}.
To investigate whether \our' representations capture such features, we applied our fully trained model to the METABRIC breast cancer cohort\citep{curtis2012genomic} further analyzed in \danenberg. As in \danenberg, we restricted the cohort to ER-positive patients ($n{=}541$).
For each patient, we mapped all cells into \our's embedding space, tiled this space into 120 phenotypic states via \(k\)-means, and distilled each tumor into a phenotype–composition fingerprint (Fig.~\ref{fig:clinical}a). Clustering these fingerprints stratified patients into two risk groups.
In UMAP visualizations, the two identified risk groups showed distinct distributions of  cell-level representations (Fig.~\ref{fig:clinical}b). Notably, when comparing the two risk groups using Kaplan-Meier survival curves, we find significantly different survival dynamics ($P < 0.001$ for a log-rank test) (Fig.~\ref{fig:clinical}c): In the high-risk group ($n{=}265$), 99 death events occurred over a 21 year horizon, whereas we only observed 58 death events in the low-risk group ($n{=}276$). 

To further assess whether the \our-based stratification reflects known multicellular structures of the patients' tumor microenvironment, we compared the distribution of risk groups against the ten structures identified in \danenberg, four of which, vascular stroma, granulocyte-enriched, suppressed expansion, and APC-enriched, were previously identified to be relevant to clinical outcome based on their hazard ratios. We calculated the risk ratio for each structure's occurrence in the high-risk group (Fig.~\ref{fig:clinical}d). The risk ratios showed strong concordance between our stratification and these known prognostic structures. Specifically, vascular stroma, linked to reduced hazard, was underrepresented in the high-risk group (risk ratio 0.399), whereas suppressed expansion and APC-enriched structures, both associated with adverse outcomes, were markedly enriched (risk ratios 4.464 and 5.636, respectively).
In contrast, a patient stratification derived solely from the survival data (without \our), by separating deceased and early censored (high-risk) individuals from surviving and late censored (low-risk) individuals, showed no significant tendency for vascular stroma or APC enriched niches to appear, and an only mildly increased risk ratio of 1.472 for suppressed expansion in the high-risk group (Suppl. Fig.~\ref{suppfig:danenberg_survival_based_groups}). 

\subsection*{Tissue-level representations support clinical diagnostics and treatment planning}
Akin to a pathologist, an effective foundation model must integrate diverse tissue features into coherent diagnostic and prognostic predictions. To assess this ability, we evaluate \our' tissue-level representation using attention-based multiple instance learning (ABMIL)\citep{ilse2018attention} on various diagnostic and prognostic classification tasks (Fig.~\ref{fig:clinical}e).

Our model's tissue-level predictions (Fig.~\ref{fig:clinical}f) showed particularly strong performance on key clinical diagnostic tasks, such as cancer sub-typing and tumor grade classification on lung cancer as well as ER status determination, grade classification on breast cancer. For lung cancer in \cords, \our' representation achieved macro-F1 scores of 0.856 for cancer subtyping and 0.530 for grading, improvements of +8.9\% and +21.8\% (both $P<0.005$) over the best baseline (\kronos), respectively. Similarly, for breast cancer from \danenberg, \our\ consistently surpassed all baselines: ER status prediction (0.806 macro-F1, +14.2\% over \kronos, $P<0.006$), ERBB2 classification (0.648, +2.0\%, $P<0.007$), grading (0.490, +15.9\%, $P<0.008$), and PAM50 subtyping (0.385, +8.2\%, $P<0.1$), despite class imbalance affecting multiple of these prediction tasks (Suppl. Fig.~\ref{suppfig:support_tissue_tasks}). 

On the NeoTRIP cohort\citep{gianni2022pathologic,wang2023spatial}, \our\ showed strong performance in predicting treatment response (pathological complete response after 24 weeks of treatment) in TNBC. Here, the detailed annotations allowed distinguishing patients who received both chemotherapy (carboplatin, nab-paclitaxel) and anti-PD-L1 immunotherapy (atezolizumab), as well as comparing performance across samples collected before, during, and after treatment (see Suppl. Fig.~\ref{suppfig:support_tissue_tasks} for class distributions). 
On \textit{pre-treatment} samples, \our\ reached a macro-averaged F1-score of 0.676, outperforming \kronos (+20.2\%, $P<0.008$), \resnet (+8.6\%, $P<0.07$), and matching \camae (0.685, $P>0.89$).
On \textit{on-treatment} samples, \our\ reached a macro-averaged F1 score of 0.714, with improvements of +30.2\%, +31.4\%, and +19.5\% over \kronos, \camae, and \resnet, respectively (all $P<0.005$).
For \textit{post-treatment} samples, \our\ reached a macro-averaged F1-score of 0.678, again exceeding \kronos (+18.63\%, $P<0.005$), \camae (+3.45\%, $P<0.30$), and \resnet (+11.77\%, $P<0.09$). These findings showcase \our' ability to capture discriminatory tissue features predictive of treatment outcomes and thus its potential utility in clinical practice.

\subsection*{Spatial and marker attention provides explainability of algorithmic decisions }
To better understand and validate how \our\ arrives at its predictions, we examined the model’s two complementary attention readouts at inference: spatial maps over the tissue and per-marker attention vectors (Suppl. Fig.~\ref{suppfig:attention_scores}). The spatial maps revealed coherent, anatomically plausible foci in both adenocarcinoma and squamous cell carcinoma sections (Suppl. Fig.~\ref{suppfig:attention_scores}a,b). Notably, for subtype classification, aligning with the regions that pathologists typically scrutinize for diagnostic assessment, attention consistently concentrated on tumor-rich epithelial compartments and on distinctive cellular neighborhoods (including tumor–immune interfaces), while stromal regions were comparatively down-weighted.

To quantify the contribution of individual markers within specific regions of interest (ROIs), we summarized the marker-attention vectors into \emph{importance scores} (see Methods for details). These scores recapitulated the dominant biology of each niche (Suppl. Fig.~\ref{suppfig:attention_scores}c). In immune-infiltrated tumor regions of squamous cell carcinoma (top ROI), the model assigned high importance to panCK and MMP11, consistent with epithelial carcinoma identity and tumor aggressiveness \citep{lu2016identifying,ma2021paradoxical}, as well as to CD45RA and CD10, which capture T- and B-cell subsets \citep{tian2017unique,louhichi2018stromal}. In fibroblast-enriched stroma with interspersed immune cells (center ROI), Vimentin (VIM), smooth muscle actin (SMA), Collagen~I, and CD248 received the highest weights, aligning with fibroblast/myofibroblast biology and extracellular matrix remodeling \citep{kalluri2006fibroblasts,tommelein2015cancer}. In immune cell–dense regions (bottom ROI), attention concentrated on CD45RA, HLA-DR, and CD20, markers of immune identity and activation, including B-cell compartments \citep{tian2017unique,louhichi2018stromal}. 

Taken together, the spatial and marker attention signals provide mutually reinforcing explanations of {\our'} decisions: the model highlights the same tumor compartments and molecular readouts that human experts rely on, and it does so in a manner that is region-specific, marker-aware, and biologically consistent with the underlying tissue ecology.

\subsection*{Retrieval via the Virtual Tissues database for clinical decision support}

To assess clinical utility, we use \our\ to retrieve clinically similar patients from a reference cohort for case-based decision support (Fig.~\ref{fig:clinical}g). This could enable clinicians to make informed decisions by comparing tissue phenotypes, applied therapies, and clinical outcomes across all previously measured cases.

\our\ can represent a patient's tissue via niche summary tokens, which are trained to capture tissue composition, architecture and interactions between different cellular neighborhoods. These tokens can thus be used to query a Virtual Tissues Database for archival cases with similar tokens.
To achieve this, we measure the similarity between \our' niche summary tokens using an optimal transport-based approach. Concretely, we compute the pairwise Wasserstein distance \citep{cuturi2013sinkhorn, peyre2019computational} between the respective niche summary tokens of the query tissue and samples in the \our\ Database. The most similar matches, along with the clinical records of the corresponding patients are retrieved and can be used for clinical decision support. 

In order to quantify \our\ ability to retrieve appropriate samples from the database, we designed three quantitative tasks and benchmark \our\ against baselines (\resnet, \kronos, \camae) and random retrieval on the primary lung cancer dataset of \cords. 
First, we compared whether the the retrieved samples matched the reference samples in their clinical and diagnostic variables (Fig.~\ref{fig:clinical}h). Across all variables, \our\ achieved significantly higher matching rates than random retrieval (validated via McNemar’s test), with particularly strong performance in matching cancer subtype, where it exceeded all baselines in mean precision.
Second, we compared cell type distributions between retrieved and reference tissues using the L1 distance between normalized histograms. Again, \our\ outperformed all baselines by achieving the lowest distributional distance (Fig.~\ref{fig:clinical}i).
Third, to evaluate similarity in molecular tissue structure, we introduced an optimal transport–based metric that computes the sliced Wasserstein distance between raw pixel values of multiplex images \citep{bonneel2015sliced}. Here, \our\ retrieved archival samples whose molecular composition was highly similar to the query (Fig.~\ref{fig:clinical}j), ranking second overall behind \kronos.
Together, these results demonstrate that retrievals based on \our' niche summary tokens capture similarities of tissues across features at multiple scales, including molecular composition, cell type composition, and clinical variables.

Three case studies depicted in Fig.~\ref{fig:clinical}k illustrate \our' ability to retrieve relevant matches across different cancer presentations:
\begin{enumerate*}[label=(\arabic*)]
    \item A grade 2 adenocarcinoma case retrieving a case with similar grade, lymph node metastasis and relapse outcomes (Fig.~\ref{fig:clinical}k, left),
    \item a grade 2 squamous cell carcinoma case matching a sample with similar tissue architecture but varying lymph node metastasis status and clinical trajectory (Fig.~\ref{fig:clinical}k, middle), and
    \item a grade 2 adenocarcinoma case matched with a previous adenocarcinoma case with comparable grade and clinical trajectory but distinct lymph node metastasis status (Fig.~\ref{fig:clinical}k, right).
\end{enumerate*}
Further case studies are exhibited in Suppl. Fig. \ref{suppfig:retrieval_examples_cords}.

\our\ retrieval prioritized cases that were both molecularly and architecturally similar to the query and enriched for the same cancer subtype, providing evidence that the learned similarity is clinically meaningful. This establishes a quantitative, case-based framework for comparing tissue phenotypes across patients.

\begin{figure*}[hbtp]
    \centering
    \includegraphics[width=\textwidth]{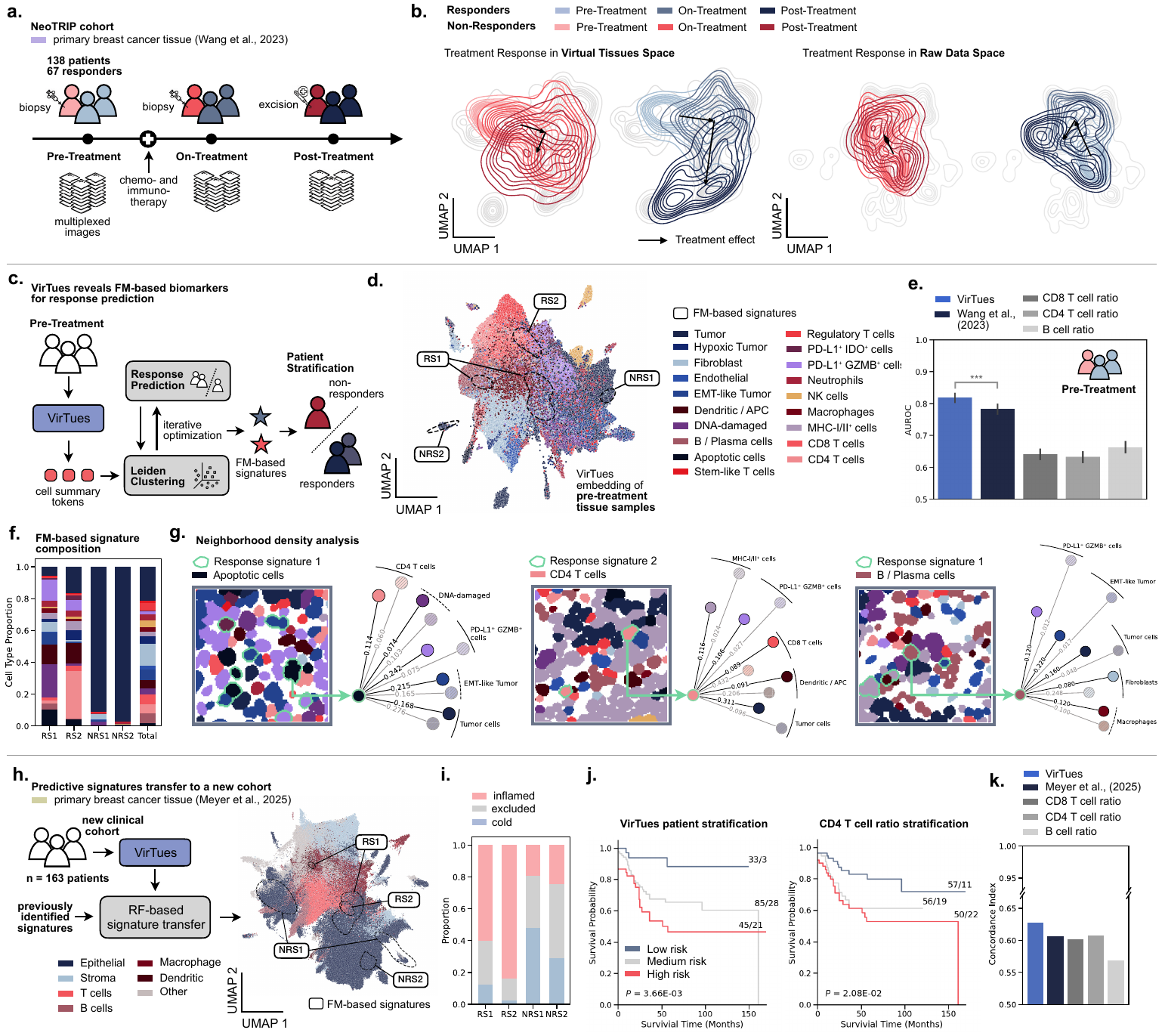}
    \caption{
    \textbf{Discovery of FM-based signatures predictive for treatment response and survival.}
    \textbf{a,} Overview of the NeoTRIP cohort\citep{gianni2022pathologic} imaged by \citet{wang2023spatial}. The study included 138 breast cancer patients who received chemo-immunotherapy, of whom 67 showed a complete pathological response. For each patient, samples were collected at up to three time points: before, during, and after treatment.
    \textbf{b,} Distributions of cell states across treatment stages, shown separately for non-responders and responders. Distributions are visualized as kernel density estimates over the UMAP embeddings of \our{}' cell-level representations (left) and mean marker abundances (right). Arrows indicate the average distribution shifts between stages.
    \textbf{c,} Cell-level representations of pre-treatment samples are iteratively clustered using Leiden at different resolutions. Predictive values of the obtained clusters are evaluated individually. Subsequently, the four most predictive clusters are selected as signatures for joint prediction of treatment response. 
    \textbf{d,} UMAP embeddings of pre-treatment cell-level representations colored by cell type. Dashed contour lines indicate distributions of selected response signatures (RS) and non-response signatures (NRS).
    \textbf{e,} Cross-validated AUROC scores for response prediction using the identified signatures on pre-treatment samples. Performance is compared with the spatial predictor system of \citet{wang2023spatial} and with three univariate baselines: the ratios of tumor cells to CD4 T cells, CD8 T cells, and B cells. Grey brackets indicate the significance of the improvement of \our\ over the second-best stratification system, as validated by an independent t-test ($P<0.001$).
    \textbf{f,} Cell type composition of the selected response and non-response signatures, along overall cell type proportions.
    \textbf{g,} Comparison of neighborhood densities of selected cell types inside and outside the response signatures (RS). Tree diagrams show, from left to right, neighborhood composition of apoptotic cells inside versus outside RS1, CD4 T cells inside versus outside RS2, and B / Plasma cells inside versus outside RS1. Black arrows indicate densities within the signature, and gray arrows indicate densities outside the signature. Only the five most frequent neighboring cell types are shown. A solid arc indicates a significant difference between inside and outside the signature ($P<0.05$ for proportions z-test), while a dashed arc indicates a non-significant difference ($P>0.05$). Cell type masks visualize representative example niches.
    \textbf{h,} \our' representations permit the transfer of each identified predictive signature using a Random Forest classifier to a new breast cancer cohort. 
    \textbf{i,} Proportions of cells with transferred signatures belonging to immune-inflamed, excluded and cold tumors.
    \textbf{j,} Risk stratification based on transferred signature proportions per patient (left) and tumor-to-CD4 T cell ratio (right). Kaplan-Meier survival curves are shown for each group, with the p-values from a log-rank test comparing the high- and low-risk groups indicated.
    \textbf{k,} Concordance index of cluster-derived risk groups compared against risk groups identified by \citet{meyer2025stratification} and three baselines, namely risk scores derived from tertile-transformed ratios of tumor cells to CD4 T cells, CD8 T cells, and B cells.
    }
    \label{fig:discovery}
\end{figure*}

\subsection*{\our\ enables the discovery of FM-based biomarkers predictive for treatment response and disease-free survival}
Motivated by our results on response prediction in the NeoTRIP TNBC cohort \citep{gianni2022pathologic, wang2023spatial} (Fig.~\ref{fig:discovery}a), we asked whether \our\ representations can yield interpretable, spatially grounded biomarkers of response to combined chemo-immunotherapy.

We first compared cell-state distributions across pre-, on-, and post-treatment samples, stratified by outcome. Using \our’s cell-level embeddings, we observed clear population-level shifts over treatment, visualized by UMAP with kernel density overlays; responders exhibited larger shifts than non-responders (Fig.~\ref{fig:discovery}b, left), consistent with therapy-induced remodeling of the tumor microenvironment. These shifts were attenuated when using raw mean marker intensities (Fig.~\ref{fig:discovery}b, right), indicating that \our\ captures treatment-relevant structure beyond first-order signal.

We then derived \emph{virtual biomarkers} from pre-treatment biopsies via an unsupervised-then-supervised pipeline (Fig.~\ref{fig:discovery}c). Specifically, we (i) embedded all pre-treatment cells with \our; (ii) performed multi-resolution Leiden clustering to obtain a hierarchy of phenotypic partitions; (iii) aggregated cells per patient to compute cluster frequencies; and (iv) scored each cluster’s predictive value for response using cross-validated, patient-level models, ranking by out-of-fold AUROC and response risk (see Methods for details). From the top hits, we retained two response-associated signatures (RS1, RS2) and two non-response signatures (NRS1, NRS2) for downstream analysis. Projected into UMAP space, these signatures occupied distinct regions and comprised mixtures of cell types rather than single canonical labels (Fig.~\ref{fig:discovery}d).

In univariate prediction, RS1, RS2, NRS1, and NRS2 achieved cross-validated AUROCs of 0.783, 0.707, 0.599, and 0.577, respectively. A multivariate logistic model combining all four signatures, and with this the resulting \our-derived biomarkers, reached a cross-validated AUROC of 0.817, surpassing the spatial predictor of \citet{wang2023spatial} by +4.53\% ($P<0.001$) and outperforming classical immune-ratio baselines commonly used in the clinic\citep{loi2020relationship,nederlof2024neoadjuvant} (tumor-to-CD8 T cell, CD4 T cell, and B cell ratios) by 23–30\% ($P<0.001$; Fig.~\ref{fig:discovery}e).

To better understand the biology captured in the \our' signatures, we analyzed the cell type composition within each signature revealing distinct immunological and phenotypic profiles (Fig.~\ref{fig:discovery}f; Suppl. Fig.~\ref{suppfig:wang_cluster_cell_type_rel_enrichment}): 
RS1 was significantly enriched in apoptotic cells ($+415.33\%$), cells showing markers of DNA damage ($+496.69\%$), and PD-L1\textsuperscript{+} GZMB\textsuperscript{+} cells ($+606.75\%$), indicating that this signature is strongly associated with stressed, damaged, immune-targeted or apoptotic cells.
In RS2, a strong enrichment of CD4\textsuperscript{+} T cells ($+409.96\%$), PD-L1\textsuperscript{+} GZMB\textsuperscript{+} cells ($+275.25\%$), and PD-L1\textsuperscript{+} IDO\textsuperscript{+} cells ($+217.74\%$) point to the involvement of active T helper cells and cytotoxic activity potentially regulated by immune checkpoints. 
In contrast, both non-response signatures were dominated by tumor cells ($+328.49\%$ and $+358.00\%$ resp.)
and showed marked depletion of all immune and stromal cell types (e.g., $-76.205\%$ fibroblasts in NRS1 and $-97.085\%$ CD4\textsuperscript{+} T cells in NRS2).

To further characterize the micro-environmental context of the selected signatures, we analyzed their cellular neighborhood compositions (Fig.~\ref{fig:discovery}g). 
In RS1, apoptotic cells were surrounded by significantly higher numbers of CD4\textsuperscript{+} T cells ($+90.00\%$) and PD-L1\textsuperscript{+} GZMB\textsuperscript{+} cells ($+222.67\%$). Moreover, neighborhoods of B/plasma cells in RS1 were strongly enriched in PD-L1\textsuperscript{+} GZMB\textsuperscript{+} cells ($+900.00\%$) and EMT-like tumor cells ($+1194.12\%$). 
Together, these observations support that RS1 reflects a highly stressed tumor microenvironment, where apoptotic and DNA-damaged cells coexist within strong cytotoxic immune activity. The enrichment of CD4\textsuperscript{+} T cells and PD-L1\textsuperscript{+} GZMB\textsuperscript{+} cells, together with EMT-like tumor cells near B/plasma cells, suggests active immune pressure on the tumor cells.
By contrast, CD4\textsuperscript{+} T cells in RS2 were located in neighborhoods with more tumor cells ($+223.96\%$), MHC-I/II\textsuperscript{+} cells ($+383.33\%$), and PD-L1\textsuperscript{+} GZMB\textsuperscript{+} cells ($+292.59\%$) but fewer CD8\textsuperscript{+} T cells ($-79.40\%$), supporting our cell type based hypothesis that RS2 captures an active tumor-proximal T helper cell response which is balanced by checkpoint regulation.

Overall, we note that response signatures 1 and 2 capture complementary anti-tumor immune mechanisms: response signature 1 reflects a cytotoxic, stress-driven response, while response signature 2 represents a T helper cell response regulated by checkpoints. Their distinct spatial and functional profiles likely explain their predictive value for anti-PD-L1 therapy, as both indicate productive immune activation which can be amplified by targeting immune checkpoints.

To assess generalizability, we transferred these signatures to an independent TNBC cohort \citep{meyer2025stratification} not used in \our\ pretraining or signature discovery. We computed \our\ cell embeddings per patient and applied random-forest classifiers trained on the discovery cohort to assign RS/NRS labels (Fig.~\ref{fig:discovery}h). As expected, transferred RSs localized to immune-inflamed regions, whereas NRSs mapped to immune-excluded/cold areas (Fig.~\ref{fig:discovery}i). Lacking immunotherapy response labels in this cohort, we evaluated prognostic utility on disease-free survival. Patient-level signature frequencies were combined into a single risk score and thresholded into three strata (Methods). Kaplan–Meier curves showed clear separation ($P<0.005$ for a log-rank test), with 3 events in low risk ($n{=}33$) versus 21 in high risk ($n{=}45$) (Fig.~\ref{fig:discovery}j). Our stratification achieved a concordance index of 0.628, exceeding the system of \citet{meyer2025stratification} (0.606) and ratio-based baselines, namely tumor-to-CD8 T cell (0.602), tumor-to-CD4 T cell (0.608), and tumor-to-B cell ratios (0.569) (Fig.~\ref{fig:discovery}k, see Suppl. Fig.~\ref{suppfig:baseline_survival_curves_meyer} for Kaplan-Meier curves).

%% file: content/discussion.tex
\section*{Discussion}\addcontentsline{toc}{section}{Discussion}

Multiplex tissue imaging promises mechanistic insight and clinically actionable biomarkers by resolving how tumor and immune cells are organized in space. Realizing this promise requires computational models that can integrate heterogeneous, high-dimensional marker panels across cohorts and institutions, support diverse discovery and prediction tasks, and remain interpretable to biologists and clinicians. Here we show that a single marker-aware foundation model trained and evaluated on diverse IMC datasets (3,102 patients; 147 markers) can unify such measurements into ``virtual tissues'' that support reconstruction, annotation, clinical prediction and cross-cohort biomarker discovery without task-specific fine-tuning.
Three design choices underpin this generality: the integration of protein language model embeddings to encode marker identity, factorized spatial/marker attention to scale to highly-multiplexed data, and hierarchical summary tokens that enable analysis across scales. Together these elements allow VirTues to operate across panels and cohorts while remaining \emph{interpretable}.

Across reconstruction tasks, VirTues learned both spatial organization and inter-marker relationships: it accurately inpainted independently masked regions, recovered fully masked markers, and restored occluded niches. In a zero-shot evaluation on an external lung cancer cohort containing markers unseen during training, the model still produced visually and quantitatively plausible reconstructions, with performance decrements largely confined to the most challenging setting of masking an entire unseen marker. This pattern indicates that VirTues relies on transferable molecular priors in addition to local spatial cues, a property that opens practical avenues for \emph{virtual augmentation} of panels and panel design \citep{jain2025test}. Such augmentation, grounded in related measured channels and biologically informed marker embeddings, may be a more reliable path than translating from information-poor stains (e.g., H\&E) to high-plex panels. 

At the cellular scale, simple linear probes on VirTues cell tokens exceeded recent encoders across multiple datasets and retained performance in zero-shot settings, including for rare cell types. Performance improved when the model was trained on a larger, more diverse corpus, a hallmark of FMs, underscoring the value of multi-dataset pretraining for spatial proteomics. Importantly, VirTues enabled \emph{label-efficient} workflows: a classifier trained on one lung cohort transferred to another with partially overlapping panels and outperformed competing representations and intensity-based baselines, particularly on clinically relevant but scarce cell classes.

VirTues also aggregated cellular information into coherent tissue-level readouts. In the ER-positive METABRIC cohort, patients clustered by phenotype–composition fingerprints derived from VirTues embeddings stratified into risk groups with distinct survival trajectories; the groups aligned with known multicellular structures linked to outcome (e.g., vascular stroma depletion and APC-enriched niches in the high-risk group), providing biological face validity. Tissue-level classifiers further delivered competitive diagnostic and prognostic performance across lung and breast cancers.

Beyond prediction, VirTues supported \emph{retrieval} of clinically similar cases using niche-level embeddings and optimal-transport distances. Retrieved neighbors matched query cases in cancer subtype and showed close cellular and molecular composition, suggesting a principled substrate for case-based discussions and hypothesis generation in molecular tumor boards \citep{vasilev2025mtbbench}. While such retrieval should be validated on prospective clinical cohorts and embedded within clinical governance, these results illustrate how FMs can bridge representation learning and evidence lookup.

A central \emph{translational} contribution is the derivation of FM-based \emph{virtual biomarkers} from pre-treatment biopsies in triple-negative breast cancer. Using a principled unsupervised–then–supervised pipeline (multi-resolution Leiden to define phenotypes followed by patient-level scoring with cross-validated AUROC/response risk), we identified two response signatures and two non-response signatures that jointly predicted immunotherapy response with AUROC 0.817, surpassing published spatial predictors and classical immune-ratio baselines. Composition and neighborhood analyses revealed complementary immune programs: a cytotoxic, stress-associated context (apoptosis/DNA damage with PD-L1\textsuperscript{+}GZMB\textsuperscript{+}) and a tumor-proximal CD4\textsuperscript{+} helper program with checkpoint features. Crucially, the same signatures transferred to an independent cohort to stratify disease-free survival and outperformed established stratification systems and ratio-based baselines, demonstrating portability of FM-derived markers across studies. Beyond their translational value, our FM-based biomarkers represent a conceptual advance in spatial biomarker discovery. In contrast to traditional approaches, which often depend on prior hypotheses and manually engineered single-cell or neighborhood features\citep{danenberg2022breast,wang2023spatial,de2024multiplex}, our biomarkers are derived through an automated, hypothesis-free optimization process, yet remain retrospectively interpretable. Hence, they provide a powerful new framework for generating biological hypotheses and guiding future discoveries.

Collectively, these findings argue for VirTues as a \emph{generalist} representation for spatial proteomics that: (i) bridges heterogeneous marker panels and cohorts through biologically informed tokenization, (ii) exhibits scaling with data diversity, (iii) retains performance under zero-shot shifts, (iv) yields interpretable signals at molecular and spatial levels through marker and spatial attention, and (v) enables discovery and transfer of mechanistically plausible biomarkers. 

Several limitations warrant emphasis. First, zero-shot reconstruction and prediction degrade when extrapolating to markers with weak biochemical relatedness to the training set; virtual augmentation should therefore be used with calibration and uncertainty reporting and anchored to measured channels whenever possible. Second, rare cell states and uncommon architectures remain challenging, likely reflecting data scarcity rather than architectural limits, highlighting the need for targeted curation of underrepresented contexts. Third, our survival analyses, while supportive, are primarily unadjusted demonstrations of signal; rigorous covariate-adjusted models, assessment of proportional hazards across several independent clinical cohorts, and prospective validation are required for clinical adoption. Fourth, attention maps provide useful but partial explanations; future work should complement them with causal or perturbational analyses. Fifth, although our data corpus spans 15 IMC cohorts across eight organ sites, broader validation across additional diseases, tissue-processing protocols, and imaging platforms is required to test universality. Finally, while spatial proteomics offers deep molecular readouts, H\&E provides scalable and robust morphology \citep{chen2024towards,jain2025test}; multimodal models that fuse these complementary signals are therefore a natural next step.

Consequently, looking forward, two directions are particularly promising. First, extending VirTues to additional spatial modalities, such as H\&E\citep{chen2024towards,xu2024whole}, IHC\citep{pati2024accelerating,lu2024visual}, technologies such as spatial transcriptomics\citep{rosen2023universal,cui2024scgpt,heimberg2025cell,pearce2025cross,chen2025visual}, metabolomics\citep{alexandrov2023spatial}, or lipidomics\citep{fusar2025lipidomic}, and to temporally resolved datasets could unify molecular scales and dynamics in a single FM, contingent on robust alignment and molecular embeddings. Second, coupling marker-aware reconstruction with generative models may enable biologically constrained virtual multiplexing and panel design\citep{pati2024accelerating,wu2025rosie,andani2025histoplexer}, accelerating hypothesis testing while reducing assay cost and tissue use\citep{jain2025test}. 

VirTues demonstrates that foundation models for spatial biology do not need to trade generality for clinical utility: through marker-aware design and hierarchical architecture, a single model can unify heterogeneous molecular assays while discovering transferable biomarkers that outperform traditional approaches, establishing a blueprint for building  foundation models that translate directly to clinical decision-making.

%% file: content/methods.tex
\section*{Methods}\addcontentsline{toc}{section}{Methods}

\subsection*{\our architecture}

Multiplexed imaging data, such as IMC, poses modality-specific challenges to the development of scalable machine learning  algorithms. The images represent high-dimensional samples, characterized by a large number of measured channels and high spatial resolutions. On the contrary, the number of samples per dataset in terms of entire images and patients is relatively small. Further, the total number as well as the combination of channels varies between datasets as studies use different marker panels. These characteristics of the data modality hinder the simple off-the-shelf application of established vision architectures. Both convolutional neural networks (CNNs) and standard Vision Transformers (ViTs) require a constant number of input channels, typically RGB, with a fixed semantic meaning. Moreover, 
in contrast to RGB channels, which combine to produce colors, multiplex channels convey distinct biological meanings and exhibit complex interrelationships.
To address the unique challenges posed by multiplex imaging data, we propose \our, an encoder-decoder model based on the ViT architecture. \our\ is designed for the efficient processing of highly-multiplexed image data accommodating varying numbers and combinations of measured markers. Furthermore, \our\ incorporates the attribution of distinct biological meaning to each measured marker. \our\ operates on tokenized image crops of size $d_c \times d_c = 128\times128$, capturing tissue niches. Restricting \our' input to such crops increases the number and diversity of pretraining samples while decreasing the dimensionality per sample.

\paragraph{Tokenization.}
To preserve the biologically distinct meaning of each channel and allow for a flexible number of channels per image, we employ a multi-channel tokenization procedure \citep{kraus2024masked, bao2023channel}. Each channel is spatially divided into patches of size $d_p \times d_p = 8\times 8$, since this approximately captures one cell per patch. Flattening each patch results in a three-dimensional grid of image tokens $\bm{x} \in \R^{M\times H \times W\times d_p^2}$, where $M$ is the number of measured markers and $H=W=d_c/d_p$ the grid height resp. width. For all $M$ markers, we retrieve from a precomputed lookup table the corresponding protein embeddings $\pi \in \R^{M\times d_{\mathrm{PLM}}}$ given by the PLM (ESM-2\cite{lin2023evolutionary} with $d_{\mathrm{PLM}} = 640$). We refer to these embeddings as marker tokens. For each channel $m$ and each grid position $(i,j)$, we project the image token $\bm{x}_{mij}$ and the corresponding marker token $\pi_m$ to the same dimension $d_{\mathrm{model}}$ using learnable linear projections, to get $\bm{x}'_{mij} \in \R^{d_{\mathrm{model}}}$ and $\pi'_m \in \R^{d_{\mathrm{model}}}$ respectively. The image and marker tokens are fused through summation, resulting in the image tokens $\widetilde{\bm{x}} \in \R^{M\times H\times W\times d_{\mathrm{model}}}$, where $\widetilde{\bm{x}}_{mij} = \bm{x}'_{mij} + \pi'_m$.
The fusion of the marker token with the image tokens serves two main purposes: (1) enabling \our\ to differentiate the channel origins of input tokens, and (2) introducing a biologically informed prior, reflecting sequence-level protein relationships, which cannot be added through other marker tokenization schemes (such as one-hot or learnable marker embeddings). We note that this is the first of many building blocks enabling \our\ to generalize across unseen markers. Further, to allow \our\ to capture an aggregated representation for each patch, we introduce an additional layer of learnable patch summary tokens $\bm{c}\in \R^{H\times W\times d_\mathrm{model}}$, one for each spatial position. Each patch summary token $\bm{c}_{ij} \in \R^{d_\mathrm{model}}$ is initialized using the same weights.

\paragraph{Masking.}
During training, a portion of the image tokens $\set{\widetilde{\bm{x}}_{mij}}$ is masked by replacing them with a special masking token $\Box \in \R^{d_\mathrm{model}}$ initialized with learnable weights. Masking is applied channel-wise by sampling a masking ratio $r_{\mathrm{masking}}$ between 60\% and 100\% and uniformly selecting the corresponding $\ceil{r_{\mathrm{masking}} H W}$ tokens to mask within the channel. We denote the resulting 3D binary mask by $\bm{M} \in \{0,1\}^{M\times H \times W}$, where the value $1$ marks masking. Masked tokens remain linked to their specific markers, which is indicated by adding the marker tokens to the masked tokens.

\paragraph{\our\ Encoder.}
The set of all non-masked image tokens $\setconstrain{\widetilde{\bm{x}}_{mij}}{\bm{M}_{mij}=0}$ and the set of patch summary tokens $\set{\bm{c}_{ij}}$ is passed as an input to the \our\ Encoder. This encoder is constructed by modifying the vision transformer's architecture \citep{dosovitskiy2021an}, to adapt it to work with varying input channels efficiently, and capture marker correlations and spatial patterns separately. In contrast to standard Vision Transformers, which use full multi-head self-attention where all tokens attend pairwise to each other, we use two specialized sparse multi-head self-attention mechanisms, marker attention and channel attention, akin to space and time attention employed in video transformers\cite{bertasius2021space}. 
In marker attention, only tokens which are placed at the same spatial grid position attend to each other, thereby capturing inter-marker dependencies and correlations. We denote the set of input tokens to the $\ell$-th transformer block as $\set{t^{\ell}_{mij}}$, where the token $t^{\ell}_{mij}$ is associated to the $m$-th channel and position $(i,j)$. In this notation, we treat the layer of patch summary tokens simply as a further channel. Then, a marker attention transformer block computes
\begin{align*}
    \forall i^*,j^*: \, \setconstrain{t^{\ell+1}_{mij}}{\substack{i=i^*\\j=j^*}} = \mathrm{MHSA}(\setconstrain{t^{\ell}_{mij}}{\substack{i=i^*\\j=j^*}}).
\end{align*}
where MHSA denotes a transformer block with standard multi-head self-attention. In contrast, in channel attention, only tokens present in the same channel attend to each other, hence capturing spatial patterns across tissue. Following the notation for marker attention, a channel attention transformer block computes as 
\begin{align*}
    \forall m^*: \, \setconstrain{t^{\ell+1}_{mij}}{m=m^*} = \mathrm{MHSA}(\setconstrain{t^{\ell}_{mij}}{m=m^*}).
\end{align*}
The \our Encoder architecture consists of a sequence of $16$ transformer blocks, which alternate between blocks that use marker and channel attention. Each of the transformer blocks uses $8$ attention heads. Spatial positions are encoded using 2D rotatory position embeddings \citep{su2024roformer}. Further, we use pre-layer normalization\citep{xiong2020layer}.

The \our\ Encoder outputs a set of encoded image tokens $\setconstrain{\widetilde{\bm{x}}^{\mathrm{enc}}_{mij}}{\bm{M}_{mij}=0}$ and a set of encoded patch summary tokens $\set{\bm{c}^{\mathrm{enc}}_{ij}}$.

\paragraph{\our\ Decoder.}
The \our\ Decoder is used during training and inference to reconstruct the original image. It is comprised of a Vision Transformer\citep{dosovitskiy2021an} followed by a single linear projection. To reconstruct the original image, the encoded tokens $\setconstrain{\widetilde{\bm{x}}^{\mathrm{enc}}_{mij}}{\bm{M}_{mij}=0}$, the encoded patch summary tokens $\set{\bm{c}^{\mathrm{enc}}_{ij}}$ and the masked tokens $\setconstrain{\widetilde{\bm{x}}_{mij}}{\bm{M}_{mij}=1}$ are regrouped as follows: For each channel $m^*$, we group the encoded and the masked tokens of that channel with a copy of the encoded patch summary tokens:
\begin{align*}
  \setconstrain{\widetilde{\bm{x}}^{\mathrm{enc}}_{m^*ij}}{\bm{M}_{m^*ij}=0} \cup \setconstrain{\widetilde{\bm{x}}_{m^*ij}}{\bm{M}_{m^*ij}=1} \cup \set{\bm{c}^{\mathrm{enc}}_{ij}}.
\end{align*}
These groups of tokens are passed individually to the decoder one by one. Hence, in the decoder tokens of different channels do not interact with each other. This design is intended to force the decoder to extract the majority of the information to reconstruct each channel from the patch summary tokens rather than relying on other channels and thus incentivize the encoder to store a meaningful representation in these. This regrouping further limits the individual token set sizes to $2HW$, thus allowing us to use full multi-head self-attention instead of marker and channel attention. Processing all tokens by the decoder's transformer followed by the linear projection yields the final grid of reconstructed image tokens $\bm{x}^{\mathrm{rec}} \in \R^{M\times H \times W\times d_p^2}$.

Following \citet{he2022masked}, we setup the encoder-decoder framework in an asymmetric fashion, where the size of encoder is deeper than the decoder, allowing the major workload of the model to rely on the encoder rather than the decoder. We construct a shallow decoder consisting of 4 transformer blocks with full attention (in contrast to 16 transformer blocks in the encoder with alternating marker and channel attention). Similar to the encoder, we use 2D rotatory position embeddings\citep{su2024roformer} to encode spatial positions and pre-layer normalization\cite{xiong2020layer} in the decoder.

\paragraph{Aggregation into cell-, niche- and tissue-level representations.}
During inference, \our\ Encoder represents each image crop as a grid of patch summary tokens $\bm{c}^{\mathrm{enc}}\in \R^{H\times W\times d_{\mathrm{model}}}$. These are aggregated into cell-, niche- and tissue-level representations as follows: For cell-level representations, the full multiplexed image is divided into a grid of overlapping crops of size $128\times128$ using as stride of $s=42$. Next, each crop is embedded independently. Finally, for each cell, the cell summary token is computed as a weighted average of all patch summary tokens, where the weight assigned to each patch equals the number of pixels intersecting with the cell in that patch. For niche- and tissue-level representations, the image is divided into a grid of non-overlapping crops of size $128\times128$, crops covered by less than 30\% of tissue are excluded, and the remaining crops are individually embedded. The niche or tissue-level representations are then obtained by aggregating all encoded patch summary tokens from the crop or the full image, respectively. For unsupervised tasks, such as the retrieval experiments (Fig. \ref{fig:clinical}g-k), a simple average $z = \frac{1}{HW}\sum_{i,j}\bm{c}_{ij}^{\mathrm{enc}}$ is used for aggregation, generating task-agnostic embeddings. In supervised settings, a dynamically weighted average is employed, achieved by training an attention-based multiple instance learning classifier \citep{ilse2018attention} on the given task (while the parameters of \our' are kept frozen), generating task-specific embeddings. This attention weighted average computes as
\begin{align*}
    z &= \sum_{i,j} a_{ij} \bm{c}_{ij}^{\mathrm{enc}}\\
    a_{ij} &= \frac{\exp{w^T(\tanh{(V\bm{c}_{ij}^{\mathrm{enc}})}\odot \sigma(U\bm{c}_{ij}^{\mathrm{enc}})})}{\sum_{i'j'}\exp{w^T(\tanh{(V\bm{c}_{i'j'}^{\mathrm{enc}})}\odot \sigma(U\bm{c}_{i'j'}^{\mathrm{enc}})})},
\end{align*}
where $U, V\in \R^{d_{\mathrm{hidden}} \times d_{\mathrm{model}}}$ and $w\in \R^{d_{\mathrm{hidden}}}$ are learnable weights, $\sigma$ is the sigmoid activation function and $\odot$ indicates element-wise multiplication. In a multi-head setting, this computation is repeated for each head with a different weight vector $w^h\in \R^{d_{\mathrm{hidden}}}$ and the resulting representations are concatenated. Per default, we use 8 heads and $d_{\mathrm{hidden}}=256$.

\paragraph{Implementation details.}

To efficiently implement marker and channel attention, we reduce these mechanisms to full attention by merging either the spatial or channel axis of the batched token tensor with the batch axis, allowing subsets of tokens that attend to each other to be treated as independent sequences. During inference without masking, this reduction leverages built-in, hardware-optimized implementations of standard self-attention. However, during training, channel-wise independent masking with varying ratios and channel dropout lead to token sequences in the marker and channel attention blocks having variable lengths. This variability poses a technical challenge because efficient built-in PyTorch attention mechanisms require uniform sequence lengths within a batch. To avoid the computational overhead of adding padding tokens, we employ a dynamic re-packaging strategy in conjunction with Flash Attention-2 (FA2)'s \cite{dao2023flashattention2} support for block-diagonal masked self-attention. Non-masked tokens within a batch are repacked into a single sequence, preserving coherent subsequences of tokens that belong to the same sample and channel or spatial position. A block-diagonal mask is generated dynamically to indicate the subsequences, specifying which tokens can attend to each other. The repacked sequence and associated mask are processed using FA2's masked self-attention implementation.

\subsection*{\our pretraining}

\paragraph{Loss function.} \our\ is trained end-to-end to reconstruct image crops in a masked auto-encoding framework\citep{he2022masked, vincent2008extracting}. Our reconstruction loss is the mean squared error between the reconstructed pixels' intensity values and the original pixels' intensity values, i.e.,
\begin{align*}
    \mathcal{L}_{\mathrm{MAE}} = \norm{\bm{x}^{\mathrm{rec}} -\bm{x}}^{2}_2.
\end{align*}
Note that this loss is computed over all pixels of both masked and non-masked tokens.

\paragraph{Data augmentation.}
Before training, we first randomly sample from each tissue image 4$N$ subimages of dimension 256 $\times$ 256, where $N$ is the number of such subimages fitting within the tissue image. Subimages which are covered by less than 30\% tissue according to the tissue segmentation mask are filtered out. During training, we subsample uniformly at random crops of size 128 $\times$ 128 from the subimages. This hierarchical two-step subsampling method approximates sampling crops uniformly at random from the whole image, while avoiding an I/O-bottleneck while training. We further apply random rotations and flips to each selected crop. Moreover, to ensure \our\ learns representations robust to varying combinations of markers, and enhance its ability to generalize to unseen datasets and markers, we sample uniformly a marker dropout ratio $r_\mathrm{dropout}$ between 0 and 25\% and exclude a corresponding number of random channels from the training sample.

\paragraph{Optimization.}
We train \our\ for 150 epochs using AdamW \citep{kingma2014adam} with an effective batch size of 512. Each epoch involves iterating once over all precomputed subimages. Weight decay is applied to all weights except biases and Layer Normalization terms, following a cosine schedule starting at 0.04. The learning rate follows a cosine decay starting at $0.0002$. Training employs automatic mixed precision with 16-bit floating point precision. Gradients are clipped to a maximum norm of 1.0.

\subsection*{Datasets for \our development}

\paragraph{Dataset curation.} To pre-train \our, we curated a collection of 14 publicly available image mass cytometry datasets, primarily focused on mapping the spatial organization of tumor micro-environments across diverse cancer types and tissue sites. These include datasets from lung cancer \citep{cords2024cancer, zhu2025spatial, rigamonti2024integrating, hu2023single, allam2022spatially, schulz2024immucanpanel1}, breast cancer \citep{danenberg2022breast, cords2023cancer, schulz2018simultaneous, wang2023spatial, schulz2024immucanpanel1}, colon\citep{schulz2024immucanpanel1}, kidney\citep{schulz2024immucanpanel1}, head\citep{schulz2024immucanpanel1}, neck\citep{schulz2024immucanpanel1}, and primary and metastatic melanoma \citep{hoch2022multiplexed, moldoveanu2022spatially}, as well as non-cancerous tissues such as tonsil and endometrium \citep{hu2023single}, and both healthy and diabetic pancreas \citep{damond2019map}. An additional 15th dataset, consisting of primary breast cancer tissue samples \cite{meyer2025stratification}, was collected after pre-training and the main analysis to assess whether the identified predictive spatial biomarkers generalize to an independent cohort. Images smaller than 256 $\times$ 256 pixels were excluded. Furthermore, we filtered out images with insufficient tissue coverage, for instance due to tearing or damage to the TMA cores, resulting in a total of 8’887 distinct images. An overview of all used IMC datasets, their tissue origin as well as their sample sizes in terms of patients, images, sampled $256\times256$ crops and annotated cells, can be found in Suppl. Table~\ref{table:datasets}. 

For downstream tasks, we utilized the segmentation masks, cell labels, and clinical annotations provided by the original studies. During quality control of the dataset from \citet{cords2024cancer}, we identified a substantial proportion of inaccurately labeled cells. To address this issue, we re-annotated the dataset using a Random Forest classifier trained on a subset of 96 images that had been manually verified to contain accurate labels. In the dataset from \citet{danenberg2022breast}, we corrected misalignment issues by re-aligning the segmentation masks and cell annotations using the provided spatial coordinates.

Each pretraining dataset was divided randomly into an 80/20 train–test split, ensuring that all samples from a given patient were contained entirely within either the training or the testing set. This split was used for the pretraining of \our and the baselines as well as for all evaluations and downstream experiments involving trainable parametric models.

For each dataset, we compiled a list of markers corresponding to the image channels. We identified the amino acid sequence for each marker and computed its ESM-2 \citep{lin2023evolutionary} embedding. For mRNA markers, we used the sequences of the encoded proteins.

Additionally, for each tissue sample, we generated an approximate binary tissue segmentation mask using Otsu thresholding followed by automatic hole filling \cite{scikit-image}. 

\paragraph{Dataset preprocessing.}
For each image, intensity values are clipped channel-wise at the 99th percentile, followed by a shifted logarithm transformation with a size factor of 1, as commonly applied to scRNA-seq count data \citep{heumos2023best}. Additionally, a Gaussian blur filter with a kernel size of 3 and unit variance is used to smooth each image. Finally, each image is standardized channel-wise \citep{sypetkowski2023rxrx1, kraus2024masked} using means and standard deviations computed over each dataset. The image-wise channel percentiles as well as the dataset-wise means and standard deviations used in this preprocessing were computed only over the tissue area as defined by the tissue segmentation masks. 

\subsection*{Evaluation}

\paragraph{\our\ model instances.}
In total, we use 14 pre-training datasets and train multiple \our\ instances on different combinations of these. The default \our\ model, which is used in most experiments unless stated otherwise, is trained on all 14 datasets. To evaluate downstream zero-shot performance (i.e., applying the model to a previously unseen dataset), we train additional instances, each excluding one of the following datasets in turn: \citet{cords2024cancer}, \citet{wang2023spatial}, \citet{hoch2022multiplexed}, \citet{danenberg2022breast}, or \citet{rigamonti2024integrating}. To demonstrate the benefits of multi-dataset training, we also train a model instance on \citet{danenberg2022breast} alone (Fig.~\ref{fig:cell_annotation}c). Finally, to study the impact of the number of channels on downstream performance (Fig.~\ref{fig:overview}e), we train \our\ on \cords with marker panels restricted to 10, 20, or 40 markers (see Suppl. Table \ref{table:subsets_marker_scaling}).

\paragraph{Comparisons and baselines.} 
We compare \our\ primarily with three deep learning baselines: (1) \resnet, (2) \camae and (3) \kronos. These baselines were selected as they are, similar to VirTues, self-supervised representation learning models recently developed for or applied to microscopy images.
We utilize a pretrained \resnet\ based on the approach described by  \citet{sorin2023single}.
Specifically, each channel is individually embedded using the ResNet50 \citep{he2016deep} architecture, pretrained on ImageNet-1K \cite{deng2009imagenet}, where each channel is duplicated thrice to match the input dimension. The resulting ResNet50 embeddings are then concatenated and projected to their 9 principal components using sparse mini-batch PCA to generate the crop representation. We remark that ResNet is a convolutional neural network that directly generates spatially aggregated niche-level representations, and is thus unable to embed patches at the cellular level. Hence, we compare against \resnet\ only for niche-level and tissue-level tasks. Furthermore, since it utilizes a pretrained network and is channel-agnostic, we also compare against it for zero-shot experiments. 
Secondly, we use a channel-agnostic masked autoencoder proposed by \citet{kraus2024masked}. This model adopts a multi-channel tokenization strategy and an encoder-decoder framework similar to \our, but with key differences: each channel is assigned a separate decoder, marker identities are not encoded in the tokenization, and full attention is utilized. These design choices restrict the model's capability to scale to a large number of channels, imposes efficiency issues and hinders the model's ability to zero-shot to unseen markers or datasets. We pretrain \camae with the same reconstruction objective and procedure described in \citet{kraus2024masked}, setting the patch size to 8 to capture information at the cellular scale. We note that we train \camae\ for each dataset separately to address scaling issues and mitigate the computational bottlenecks caused by the unequal representation of channels in datasets, which would otherwise lead to a disproportionate increase in model parameters without a corresponding increase in data. From Fig.~\ref{fig:overview}e, we notice that the number of parameters in \our\ is already 25$\times$ less than \camae\ for 40 markers. \camae\ can be used to generate both patch-level and niche-level representations. For patch-level representations, we average the embedded tokens along the channel dimension. For self-supervised niche-level representations, we take the average of all embedded tokens of the crop. 
Thirdly, we compare against \kronos, a recent foundation model for spatial proteomics. For this re-train \kronos on our collection of pre-training datasets. The pre-training of \kronos requires each training sample to contain a nucleus marker. As such we use Histone H3 and exclude three pre-training datasets not measuring consistently this marker, namely \citet{zhu2025spatial}, \citet{rigamonti2024integrating} and \citet{hu2023single}. We follow the original training protocol of \kronos with one modification. \kronos is applied to fluorescence-based imaging with resolutions between 0.37-0.5\textmu m/px, using a crop size of $256\times256$ pixels. In contrast, IMC data has a resolution of 1\textmu m/px. To ensure equivalent physical field of views, we configure \kronos with a crop size of $128\times128$, a local view size of $48\times48$ and a patch size of $8\times8$ pixels. These parameters further match those of \our, thereby enabling a direct comparison between representations obtained for the same visual input. For patch-level representations, we average the token-specific features of \kronos per spatial position, akin to \camae. For the niche-level representation, we concatenate the marker-specific features and project the resulting embeddings to their leading 256 principal components, following the same procedure as used by \citet{shaban2025foundation} for patient stratification.

\paragraph{Masked reconstructions.}
We evaluate \our' understanding of molecular tissue structure and biological relationships between markers by assessing the reconstruction ability of \our\ for three different masking strategies: independent masking, marker masking and niche masking. For independent masking (Fig.~\ref{fig:masking}a and Suppl. Figs.~\ref{suppfig:examples_independent_1}, \ref{suppfig:examples_independent_2} and \ref{suppfig:examples_independent_3}),  we sample a masking ratio independently for each channel in the input image, uniformly between 60\% and 100\% and mask the corresponding number of patches. This strategy allows the model to leverage both spatial patterns and marker relationships to reconstruct the masked regions, aligning with the masking used during training. For marker masking (Fig.~\ref{fig:masking}b and Suppl. Figs.~\ref{suppfig:examples_marker_1}, \ref{suppfig:examples_marker_2} and \ref{suppfig:examples_marker_3}), we select a single marker from the input image and mask all corresponding patches. This approach enables us to evaluate \our' understanding of marker correlations in isolation of its spatial understanding. In reverse, niche masking (Fig.~\ref{fig:masking}c and Suppl. Figs.~\ref{suppfig:examples_niche_1}, \ref{suppfig:examples_niche_2} and \ref{suppfig:examples_niche_3}) is designed to analyze \our' understanding of spatial structures. For each image, we sample a single masking ratio uniformly between 60\% and 100\% and use this ratio to select the corresponding number of grid positions, where we mask all patches across all channels. We emphasize that in all three masking strategies, channel dropout is not applied during the evaluation phase.
To quantify the reconstruction ability, we compute the Pearson correlation between reconstruction and ground truth images on the test split, per dataset, marker and masking strategy (Fig.~\ref{fig:masking}d and Suppl. Fig.~\ref{suppfig:reconstruction_losses}). We further compute the pixel-wise F1-scores of predicted protein positivity, where ground-truth positivity is defined by thresholding at the mean intensity of each marker (Suppl. Fig.~\ref{suppfig:reconstruction_f1_scores}). In contrast to the training loss, we compute these metrics only for the masked tokens' pixels to ensure comparability across masking strategies, despite varying relative masking ratios. As a baseline for independent and niche masking, we calculate the performance obtained when masked pixels in each channel are inpainted using the average intensity of the visible pixels in that channel. Further, we report for each dataset and marker, the performance reached by predicting under marker masking for each channel the most highly correlated alternative marker.

\paragraph{Cell-level tasks.}
Previous work on IMC image-based learning \citep{sorin2023single} fails to decode cell types at a cellular scale. In contrast, \our' cell summary tokens capture the tissue at the scale of individual cells. We test \our' ability to capture biologically meaningful signals at this cell scale through cell phenotype classification experiments. This evaluation covers four datasets, \cords, \danenberg, \hoch, and \citet{wang2023spatial}, with two levels of class granularity considered for \cords and \citet{wang2023spatial}.
Further, to assess the robustness of the representations, we probe \our' ability to transfer cell phenotypes learnt from one labeled dataset to another dataset with only partially overlapping marker panel. For this, we utilize as a source dataset \citet{cords2024cancer} and as a target dataset \citet{rigamonti2024integrating}, both measuring non-small cell lung cancer.

For the phenotype classification task, we perform linear probing using a logistic regression model with an L-BGFS solver and L2-regularization with coefficient $\lambda=1.0$. This ensures the evaluation focuses on the quality of the learned representations rather than the complexity or configuration of the classifier. The linear probe is applied to all cell-summary tokens of the respective dataset, using the same patient-level train-test split as used during pretraining. We avoid resampling strategies as applying them solely to the training set did not yield noticeable performance improvements. The cell labels are taken from the originally published datasets, where they were typically assigned through an expert-guided combination of gating, clustering, and machine learning methods applied to the tabular single-cell data. For \danenberg, we consolidate the highly nuanced phenotypes provided by the authors into eight broad categories. Similarly, we group the phenotypes from \hoch into six broad categories. For \citet{wang2023spatial}, we define six high-level and 19 corresponding low-level categories. The mappings to regroup the phenotypes can be found in Suppl. Tables \ref{table:mapping_celltype_danenberg}, \ref{table:mapping_celltype_hoch} and \ref{table:mapping_celltype_wang}. For \cords, we used the provided groupings of six high-level and 22 low-level categories.
In summary, the tasks and respective labels are as follows:
\begin{itemize}[]
    \item \textit{\citet{cords2024cancer} (cell type, coarse)}: 6 classes, namely tumor, fibroblast, immune, T cell, vessel and other.
    \item \textit{\citet{cords2024cancer} (cell type, fine-grained)}: 22 classes, namely B cell, blood, CD4, CD8, collagen CAF, HEV, hypoxic tumor, IDO CAF, lymphatic, myeloid, neutrophil, normal tumor, PDPN CAF, SMA CAF, dCAF, hypoxic CAF, hypoxic tpCAF, iCAF, mCAF, tpCAF, vCAF and other.
    \item \textit{\citet{danenberg2022breast} (cell type)}: 8 classes, namely NK, B cell, T cell, myeloid, ER+, ER-, stromal and APC.
    \item \textit{\citet{hoch2022multiplexed} (cell type)}: 6 classes, namely tumor, lymphocytes, macrophages, stroma, T cell and other.
    \item \textit{\citet{wang2023spatial} (cell type, coarse)}: 6 classes, namely stroma, immune, tumor, T cell, vessel and other.
    \item \textit{\citet{wang2023spatial} (cell type, fine-grained)}: 19 classes, namely fibroblasts, dendritic / APC, macrophage, B / plasma cell, NK, neutrophils, tumor, hypoxic tumor, EMT-like tumor, DNA-damaged cell, apoptotic cell, CD4 T cell, CD8 T cell, regulatory T cell, stem-like T cell, endothelial, PD-L1\textsuperscript{+}GZMB\textsuperscript{+} cell, PD-L1\textsuperscript{+}IDO\textsuperscript{+} cell and MHC-I\&II\textsuperscript{+} cell.
\end{itemize}

We report the F1-score per class as well as the macro-averaged F1-score and benchmark against the two baselines, \camae and \kronos.

For the transfer experiment, we train a Random Forest classifier with 100 trees, Gini splitting criterion, bootstrapping, and a maximum depth of 16 on the train-split cell-level representations from the \cords dataset with coarse cell type annotations. Unlike the supervised classification experiments, we balance the training data via downsampling to address potential distribution shifts between datasets. We then apply this classifier to predict phenotypes for cells in \citet{rigamonti2024integrating}. To compute cell-level representations, we use only the markers common to both panels. For \citet{rigamonti2024integrating}, we derive cell annotations for the same six cell categories used to train the classifier on \cords by regrouping the provided cell phenotypes. The mapping is detailed in Suppl. Table \ref{table:mapping_celltype_rigamonti}. We evaluate our predictions against these annotations as ground truth and report per-class F1-scores for \our, \kronos, \camae (trained on \citet{cords2024cancer}) and mean marker abundances.

\paragraph{Tissue structure-based risk stratification.}
The \danenberg dataset, derived from the METABRIC study\citep{curtis2012genomic}, provides detailed survival data that enables us to assess the ability of \our\ to capture multicellular structural differences in tissues relevant to patient outcomes. Following \danenberg, we restrict the dataset to ER-positive cases, resulting in 541 tissue images. For each image, we compute cell summary tokens, which are partitioned into 120 clusters using k-means. Each tissue is then represented by a 120-dimensional cluster proportion vector. Based on these proportion vectors, we group tissues into 4 groups, again using k-means. Survival analysis reveals that two groups correspond to higher-risk patients and two to lower-risk patients; we merge them accordingly into high-risk and low-risk groups. More generally, the number of groups in this procedure can be tuned to balance the relative influence of structural features and survival data in the stratification process. 
We compute Kaplan–Meier survival curves for the resulting groups and assess their separation using a log-rank test, reporting the corresponding p-value. To connect our stratification with known biology, we leverage the 10 multicellular TME structures described by \danenberg, four of which are associated with a significantly increased or decreased hazard ratio. Specifically, we calculate the risk ratio and 95\% confidence interval for the occurrence of each structure in the high-risk group. For comparison, we construct an alternative stratification based solely on survival outcomes, à priori independent of tissue structure. In this baseline, the high-risk group consists of patients who died or were censored early, while the low-risk group includes those who survived or were censored late (Suppl. Fig.~\ref{suppfig:danenberg_survival_based_groups}a). Under this survival-based stratification, we compute also the risk ratio for the occurrence of each TME structure (Suppl. Fig.~\ref{suppfig:danenberg_survival_based_groups}b). 

\paragraph{Tissue-level classification.}
To evaluate the ability of \our' representations to capture clinically meaningful information about the tissue and the patient, we define and benchmark various tissue-level classification tasks.  We derive these tasks from the clinical patient-level annotations available in the metadata of all the published datasets, where each image is treated as an independent sample. For each task, we exclude images without labels and omit classes with ambiguous or missing descriptions. We describe the task, their sample sizes and their labels below:
\begin{itemize}
    \item \textit{\citet{cords2024cancer} (primary lung cancer tissue, $n=1969$ images)}:
    \begin{enumerate}
        \item Cancer subtype: Adenocarcinoma, Squamous cell carcinoma (restricted to these as they account for 95\% of samples)
        \item Relapse: Relapse, No relapse
        \item Grade: Grade 1, Grade 2, Grade 3
    \end{enumerate}
    \item \textit{\citet{danenberg2022breast} (primary breast cancer tissue, $n=688$ images)}:
    \begin{enumerate}
        \item ERBB2 status: ERBB2 positive, ERBB2 negative
        \item ER status: ER positive, ER negative
        \item PAM50 subtype: Normal-like, Basal, HER2, Luminal A, Luminal B
        \item Grade: Grade 1, Grade 2, Grade 3
    \end{enumerate}
    \item \textit{\citet{wang2023spatial} (primary breast cancer tissue, $n=817$ images)}:
    \begin{enumerate}
        \item Response: Pathological complete response, Residual disease
    \end{enumerate}
    Note: We restrict the analysis to the same subset of per-protocol patients as used by \citet{wang2023spatial} who received both chemotherapy and immunotherapy, and conditioned on samples collected either before ($n = 356$ images), during ($n = 228$ images), or after treatment ($n = 233$ images).
\end{itemize}

For all tasks, we use gated attention-based multiple instance learning models \citep{ilse2018attention} (Gated ABMIL), where the input is the set of patch-level representations belonging to the tissue. The ABMIL classifiers aggregate these representations using gated attention computed over eight heads, each with a hidden dimension of 256. The aggregated tissue representation is then projected to class logits, followed by a softmax to obtain class probabilities. We train the models using the Adam optimizer \citep{kingma2014adam} with a learning rate of $10^{-4}$, a batch size of 16 images, and a maximum of 100 epochs. Early stopping with a patience of 10 epochs is applied based on a validation loss computed over a 20\% hold-out validation subset of the train split. Performance is reported as the macro-averaged F1-score for each task averaged over five runs with different random seeds, initializations and validation sets. As baselines, we compare against \kronos, \camae, and \resnet. For \resnet, we use niche-level rather than patch-level representations. For comparisons, p-values are indicated using one-sided Mann-Whitney U tests performed over the five runs.

\paragraph{Quantifying treatment responses.}
We use the dataset of \citet{wang2023spatial}, which contains imaging mass cytometry (IMC) data from a cohort of breast cancer patients sampled before, during, and after treatment. For many patients, samples are available at all three time points. This dataset enables us to evaluate \our' capability to quantify treatment responses and to demonstrate the discovery of virtual spatial biomarkers predictive of response.

For our analysis, we restrict the cohort to patients who receive both chemotherapy and immunotherapy, yielding a total of 138 patients. For each patient, we compute cell summary tokens from all available samples. To quantify patient-wise the strength of treatment response between two time points $t_1, t_2$, we compute the entropy-regularized 2-Wasserstein distance between cell-level representations $\bm{a} \in \R^{N_1\times d_{\mathrm{model}}}, \bm{b} \in \R^{N_2\times d_{\mathrm{model}}}$ for those time points, i.e.,
\begin{align*}
    W_2^\varepsilon(\bm{a},\bm{b}) &= \left(\min_{\pi \in \Gamma} \sum_{i=1}^{N_1}\sum_{j=1}^{N_2}\pi_{ij}(\norm{\bm{a}_i-\bm{b}_j}^2_2 -\varepsilon\log{\pi_{ij}}) \right)^{\frac{1}{2}}
\end{align*}
with $\Gamma = \setconstrain{\pi \in \R^{N_1\times N_2}}{\pi \vec{1} = \vec{1}/N_1 \text{ and } \pi^T\vec{1}\ = \vec{1}/N_2}$. As regularization strength, we use $\varepsilon = 10^{-3}$. When for a patient, multiple tissue samples at the same time point are available, we use the union of all cell representations. To qualitatively compare the average response strength between responders and non-responders, we first sub-sample the set of all cell summary tokens by a factor of 10. We then compute two-dimensional UMAP embeddings of the sub-sampled tokens. For each patient and time point, we interpret the embedded tokens as a discrete empirical distribution. Within each cohort $C \in \{\text{responders}, \text{non-responders}\}$ and for each time point $t$, we compute the Wasserstein barycenter of the patient-level distributions $\{\bm{a}_{i,t} : i \in C\}$ with a fixed support size of $m = 500$ points. Formally, the barycenter $\bm{b}_{C,t} \in \mathbb{R}^{m \times 2}$ is obtained as
\begin{align*}
    \bm{b}_{C,t} = \arg\min_{\bm{b} \in \mathbb{R}^{m \times 2}} \sum_{i \in C} W^0_2\big(\bm{b}, \bm{a}_{i,t}\big).
\end{align*}
To visualize temporal changes between time points $t_1,t_2$ in treatment response, we represent the trajectory of each group $C$ in the UMAP embedding space as the sequence of displacement vectors between the medians
\begin{align*}
    \bm{\Delta}_{C,t} = \mathrm{med}(\bm{b}_{C,t_2}) - \mathrm{med}(\bm{b}_{C,t_1}).
\end{align*}
These vectors reflect the average shift in cell-state distribution between consecutive time points.

\paragraph{Identification of FM-based biomarkers predictive of therapy response.}
For the discovery of virtual biomarkers predictive of immunotherapy response in the \citet{wang2023spatial} dataset, we consider only patient samples collected prior to treatment. To ensure comparability with the prediction performance reported by \citet{wang2023spatial}, we restrict the analysis further to the same subset of per-protocol patients as used by \citet{wang2023spatial}, resulting in total of 111 distinct individuals. For each patient, we compute the set of all cell summary tokens merged across their available samples. 
To identify predictive FM-based biomarkers, we apply Leiden clustering iteratively, varying the resolution parameter $r$ over the interval $[4,5]$ in steps of $\Delta r = 0.05$. For each clustering, we retain only clusters containing less than 2000 cells. For every patient–cluster pair, we compute the proportion of cells belonging to that cluster. These proportions are then discretized into four ranks: rank 0 for absence (0\% occurrence) and ranks 1–3 corresponding to the tertiles of positive proportions. 
To evaluate the potential of each cluster to predict response, we use the cluster rank as the predictor in a univariate logistic regression. The performance is quantified as the mean AUROC obtained from stratified 4-fold cross-validation repeated 10 times. For each cluster, we also calculate the risk ratio of treatment response conditional on high cluster presence (i.e., rank $>1$). Clusters with a risk ratio greater than 1 are classified as response clusters, while those with a ratio less than 1 are classified as non-response clusters. 
From all clusters identified across resolutions, we select the two response clusters and the two non-response clusters with the highest individual AUROC scores as the predictive virtual spatial biomarkers. We evaluate their joint predictive performance using a multivariate logistic regression model with the concatenated cluster ranks as predictors. Performance is measured as the mean AUROC from stratified 4-fold cross-validation repeated 25 times. We compare these results with the similarly cross-validated AUROC scores reported for the spatial predictor system developed by \citet{wang2023spatial}, as well as with three baselines that use, as a univariate predictor, the ratio of tumor cells to CD4\textsuperscript{+} T cells, CD8\textsuperscript{+} T cells, or B cells. For the comparisons, we compute p-values using a one-sided independent t-test.

For further interpretation of the identified predictive clusters, we calculate their cell type composition (Fig.~\ref{fig:discovery}f) as well as relative changes in cell type proportions (Suppl. Fig.~\ref{suppfig:wang_cluster_cell_type_rel_enrichment}). In addition, for each cell type and cluster, we compute the proportion of neighboring cell types, separately for cells that belong to the cluster and for those that do not (Fig.~\ref{fig:discovery}g).

\paragraph{Transferring predictive virtual spatial biomarkers across datasets.}
We evaluate the robustness and translational potential of the identified virtual spatial biomarkers by transferring them to the independent breast cancer cohort from \citet{meyer2025stratification}. Notably, the images in this cohort were not used for pretraining \our\ or for the initial identification of clusters.

For the transfer, we train a Random Forest classifier for each selected cluster using cell-level representations. Each classifier consists of 200 trees, employs the Gini splitting criterion, bootstrapping, and is limited to a maximum depth of 16. The task for each classifier is to predict whether a given cell belongs to the respective cluster. To train each classifier, we balance the dataset by down-sampling the majority class to ensure adequate recall for the target cluster, and we hold-out 20\% of the cells as a validation set to evaluate classifier performance.
Using these trained classifiers, we predict cluster memberships for all cells in the \citet{meyer2025stratification} dataset based on their cell-level representations. For interpretation of these transferred clusters, we compute, for each cluster, the proportions of cells belonging to immune inflamed, excluded or cold tumors as labeled by \citet{meyer2025stratification}.

In contrast to \citet{wang2023spatial}, the data of \citet{meyer2025stratification} lacks immunotherapy response annotations but includes survival data. Consequently, we substitute response analysis with an evaluation of the predictive value of our cluster ranks on disease-free survival. Specifically, mirroring the original discovery procedure, we calculate, for each cluster-patient pair, the proportion of cells assigned to that cluster and discretize these proportions into four ordinal ranks. Subsequently, we compute for each patient an overall risk score $R$ defined as the sum of response cluster ranks minus the sum of non-response cluster ranks. Based on their risk score, patients are stratified into high-risk ($R>2$), medium-risk ($-1<R<2$), and low-risk ($R<-1$) groups. We calculate Kaplan-Meier survival curves for these risk groups. To validate the statistical significance of the difference between high-risk and low-risk survival curves, we compute the p-value of a log-rank test. Further, we report the concordance index of these risk groups and compare our results against four baselines, namely risk groups identified by \citet{meyer2025stratification} and risk scores derived from tertile-transformed ratios of tumor cells to CD4\textsuperscript{+} T cells, CD8\textsuperscript{+} T cells, and B cells.

\paragraph{Information retrieval.}
We construct the Virtual Tissue database for \cords by extracting the central $4\times 4$ grid of crops with each crop sized $128\times 128$, for each image, excluding those smaller than this grid. This choice of grid size maximizes the tissue area captured per image while minimizing the number of excluded images. Additionally, it often eliminates empty or irrelevant corners as a side effect. Similar to tissue-level tasks, we keep those images associated with Adenocarcinoma or Squamous cell carcinoma. The remaining crops are embedded using \our or one of the baselines and the resulting self-supervised niche-level representations are stored in the database.

The database is used to retrieve tissues similar to a given reference image, measured using the 2-Wasserstein distance between sets of niche-level representations. Given the niche-level representations $\bm{a},\bm{b} \in \R^{N\times d_{\mathrm{model}}}$ for two tissues, the 2-Wasserstein distance computes as
\begin{align*}
    W_2(\bm{a},\bm{b}) &= \left(\min_{\pi \in \Gamma} \sum_{i,j=1}^{N}\pi_{ij}\norm{\bm{a}_i-\bm{b}_j}^2_2\right)^{\frac{1}{2}},
\end{align*}
where $\Gamma = \setconstrain{\pi \in \R^{N\times N}}{\pi \vec{1} = \vec{1}/N \text{ and } \pi^T\vec{1}\ = \vec{1}/N}$.

For a given reference image, we identify the closest matches based on this distance metric.

To evaluate the efficacy of the retrieval mechanism, we perform a quantitative comparison across different embedding methods. Specifically, we compute two complementary distance metrics between each reference image and its closest retrieved match, and compare their average values against those from randomized retrieval. The first metric captures cell type composition by calculating the proportions of coarse cell types within each image (using published annotations) and measuring the L1 distance between the resulting proportion vectors. The second metric reflects molecular tissue composition by representing each image as a set of pixel vectors and computing the sliced Wasserstein distance\citep{bonneel2015sliced} between these sets. Formally, for two tissue images with pixels $x,y \in \R^{P\times C}$, where $P$ denotes the number of pixels per image, the sliced Wasserstein distance is defined as
\begin{align*}
SW_2(x,y) = \left( \int_{\mathbb{S}^{C-1}} W_2\!\left(\mathrm{proj}_{\theta}(x), \mathrm{proj}_{\theta}(y)\right)^2 \, d\theta \right)^{\tfrac{1}{2}},
\end{align*}
where $\mathrm{proj}_{\theta}$ denotes the projection of each row onto the unit vector $\theta$. We employ the sliced Wasserstein distance as a computationally efficient approximation of the Wasserstein distance, given the very large number of pixels per image.

We further evaluate clinical feature matches using the McNemar test. For each clinical feature, we compare the number of correct matches among the top three results retrieved by our Wasserstein-based method against those from three randomized retrievals, and report the corresponding p-values. Let $n_1$ denote the number of cases correctly matched by the Wasserstein-based retrieval but not by the random retrievals, and $n_2$ the number of cases correctly matched by the random retrievals but not by the Wasserstein-based retrieval. The McNemar test statistic is then defined as $\chi_0^2 = (n_1 - n_2)^2/(n_1+n_2)$ which follows a chi-square distribution with 1 degree of freedom. The p-value is given by $\Pr(\chi^2 \geq \chi_0^2)$, with $p < 0.05$ indicating a statistically significant difference in clinical label matching between the two retrieval methods.

\paragraph{Zero-shot inference.}
To evaluate \our' ability to generalize to new datasets unseen during pretraining, i.e., to perform zero-shot inference, we adopt a leave-one-dataset-out approach. Specifically, we train separate instances of \our on all pretraining datasets except one, which is withheld.  Each resulting model is then evaluated on the excluded dataset in downstream experiments. We consider two types of downstream tasks: (1) cell type classification on \citet{cords2024cancer} (coarse labels), \citet{hoch2022multiplexed}, \citet{danenberg2022breast}, and \citet{wang2023spatial} (coarse labels); and (2) masked image reconstruction on \citet{rigamonti2024integrating} using independent, marker, and niche masking. For the reconstruction task, we further distinguish between markers already encountered during pretraining in other datasets and entirely novel markers. For both task types, we adopt the same setup and evaluation protocol as in the preceding non–zero-shot experiments. Finally, we compare the performance of the zero-shot model instances against that of the model trained on all datasets.

\paragraph{Scaling analysis.}
To analyze the impact of the number of measured markers on both computational costs and prediction performance, we select nested subsets of 10, 20 and 40 markers from the original full panel of \cords, based on their presumed informativeness regarding general tissue morphology and cell type differentiation. This selection is guided by prior knowledge and domain expertise. However, we acknowledge that this process is inherently subjective, as a quantitative framework to objectively rank markers by informativeness does not exists. For a full list of markers per experiment, see Suppl. Table \ref{table:subsets_marker_scaling}. We train instances of \our and \camae on these chosen subsets on the \cords\ dataset, and report the inference computational cost, number of parameters, and downstream performance upon scaling the number of channels (Fig.~\ref{fig:overview}e). We measure the computational cost $c = m \times t$, where $m$ is the memory utilized during forward pass of a batch of 16 images, and $t$ is the inference time for the batch. We allow 10 warmup runs to remove GPU startup effects, and report the average of 100 iterations. We further report the downstream macro-averaged F1 scores achieved by linear probes for cell type classification using both coarse and fine-grained classes on \cords.
We further evaluated the effects of single-dataset versus multi-dataset pretraining by training a model exclusively on primary breast cancer tissues~\citep{danenberg2022breast}, and assessing its cell type classification performance using class-specific F1-scores (Fig.~\ref{fig:cell_annotation}c).

\subsection*{Model inspection and visualization}
We investigate \our' interpretability in learning meaningful signals by evaluating the attention scores from the marker and channel attention layers in the encoder. 
We begin by examining the attention weights learned by the first marker attention layer for an input image. We consider the post-softmax attention weights for all channels except the patch summary token, say $\alpha_{h,m',m,i,j}$ when marker $m'$ attends to marker $m$ for head $h$ and spatial position $(i,j)$, and aggregate them across all heads and spatial positions. The final scores, termed as importance scores $\mathcal{I}_m$, can be written as
\[
\begin{split}
\rho_{m,m',h} &= \sum_{i,j} \alpha_{h,m',m,i,j} \\
  \mathcal{I}_m &= \sum_{m', h} \frac{ \rho_{m,m',h} - \min_{m'}\rho_{m,m',h}}{\max_{m'}\rho_{m,m',h} - \min_{m'}\rho_{m,m',h}}.
\end{split}
\]

Visualization of spatial attention typically relies on a [CLS] token. Since the pretrained \our\ inherently does not have a [CLS] token, we augment \our\ with learnable channel summary tokens (one for each channel, including the patch summary token channel), which are spatially placed at the center of the input image. We finetune the augmented model on the cancer subtype prediction task on \cords, using the encoded channel summary token of the cell summary layer to predict the cancer subtype. 
To visualize attention maps, we compute the attention scores directed from the channel summary token to the patch summary tokens in the penultimate spatial attention layer and display these as a heatmap over the spatial positions.

\subsection*{Computing hardware and software.}

We used Python (v3.12.9) together with PyTorch (v2.5.1, CUDA 12.1), and FlashAttention-2 \citep{dao2022flashattention, dao2023flashattention2} (v2.7.4) as deep-learning frameworks. For downstream experiments and statistical analysis, we used scikit-learn \citep{scikit-learn} (v1.5.2), scikit-survival \citep{scikit-surv} (v0.24.1), lifelines \citep{davidson2019lifelines} (v0.30.0) and cuML \cite{raschka2020cuml} (v25.8.0). Model pretraining was performed on a HPC system with NVIDIA GH200 GPUs. All downstream experiments were executed on NVIDIA A100 80 GB GPUs.